\pdfoutput = 1

\documentclass[a4paper,12pt]{article}
\usepackage{hyperref}

\usepackage{graphicx}
\usepackage{caption}
\usepackage{cite}
\usepackage{color}
\usepackage{subfig}
\usepackage{float}
\usepackage{slashed,dsfont}
\usepackage{longtable}
\usepackage{booktabs}
\usepackage{amsmath}
\usepackage{amssymb}

\usepackage{tabularx}
\usepackage{rotating}
\usepackage{lscape} 
\usepackage{epsfig}
\usepackage{color}
\usepackage{array}

\restylefloat{figure}

\usepackage[english]{babel}

\usepackage{geometry}
\numberwithin{equation}{section}
\newcommand{\nn}{\nonumber}

\def\mLb{{m_{\Lambda_b}}}
\def\mmLb{{m^2_{\Lambda_b}}}
\def\mL{{m_{\Lambda}}}
\def\mmL{{m^2_{\Lambda}}}
\def\cl{{\cos\theta_\ell}}
\def\ccl{{\cos^2\theta_\ell}}
\def\plpl{{+\frac{1}{2}+\frac{1}{2}}}
\def\plmi{{+\frac{1}{2}-\frac{1}{2}}}
\def\mipl{{-\frac{1}{2}+\frac{1}{2}}}
\def\mimi{{-\frac{1}{2}-\frac{1}{2}}}
\def\re{{\rm Re}}
\def\mC{{\mathcal{C}}}

\definecolor{schrift}{RGB}{120,0,0}

\begin{document}
\begin{titlepage}

\vspace{1.2cm}
\begin{center}
{\Large\bf\color{schrift} Model independent New Physics analysis in $\Lambda_b\to\Lambda\mu^+\mu^-$ decay}
\end{center}
		
\vspace{0.5cm}
\begin{center}
{\sc Diganta Das } \\[0.3cm]
{\sf Department of Physics and Astrophysics, University of Delhi, Delhi 110007, India}
\end{center}
		
\vspace{0.8cm}
\begin{abstract}
\vspace{0.2cm}\noindent

\noindent We study the rare $\Lambda_b\to\Lambda\mu^+\mu^-$ decay in the Standard Model and beyond. Beyond the Standard Model we include new vector and axial-vector operators, scalar and pseudo-scalar operators, and tensor operators in the effective Hamiltonian. Working in the helicity basis and using appropriate parametrization of the $\Lambda_b \to \Lambda$ hadronic matrix elements, we give expressions of hadronic and leptonic helicity amplitudes and derive expression of double differential branching ratio with respect to dilepton invariant mass squared and cosine of lepton angle. Appropriately integrating the differential branching ratio over the lepton angle, we obtain the longitudinal polarization fraction and the leptonic forward-backward asymmetry and sequentially study the observables in the presence of the new couplings. To analyze the implications of the new vector and axial-vector couplings, we follow the current global fits to $b\to s\mu^+\mu^-$ data. While the impacts of scalar couplings can be significant, exclusive $\bar{B}\to X_s\mu^+\mu^-$ data imply stringent constraints on the tensor couplings and hence the effects on $\Lambda_b\to\Lambda\mu^+\mu^-$ are negligible.

\end{abstract}

\end{titlepage}

\newpage
\pagenumbering{arabic}


\section{Introduction}
The Standard Model (SM) of particle physics has so far been enormously successful in explaining most of the particle physics experiments. Recently however, discrepancies between SM predictions and experimental measurements have been observed in several decay modes of $B$ mesons. For example, LHCb measurements \cite{Aaij:2014ora,Aaij:2017vbb} of the ratio of branching ratios of $\bar{B}\to K(K^\ast)\ell^+\ell^-$ decays into di-muons over di-electrons deviate from the SM predictions $\sim 1$ by about 3$\sigma$. Other notable deviations in the $b\to s\mu^+\mu^-$ mediated transitions include the so called $P_5^\prime$ anomaly between theory \cite{Descotes-Genon:2013wba} and experiments \cite{DescotesGenon:2012zf,Aaij:2015oid,Abdesselam:2016llu,Wehle:2016yoi} in the $\bar{B}\to K^\ast\mu^+\mu^-$, and the systematic deficit in the branching ratio of $\bar{B}_s\to\phi\mu^+\mu^-$ \cite{Aaij:2013aln,Aaij:2015esa}. In the $b\to c\ell\nu$ mediated transitions, ratio of $B\to D^{(\ast)}\ell\nu$ decay involving $\tau$ leptons over light leptons has been measured by BaBar \cite{Lees:2012xj,Lees:2013uzd}, Belle \cite{Sato:2016svk,Huschle:2015rga,Hirose:2016wfn} and LHCb \cite{Aaij:2015yra,Aaij:2017uff} which deviate from the SM predictions \cite{Na:2015kha,Fajfer:2012vx} by about 2-3$\sigma$ with a combined deviation of $\approx 4\sigma$. Similar deviations from the SM prediction in $B^+\to J/\psi\ell^+\nu$ decay \cite{Wen-Fei:2013uea,Dutta:2017xmj} has also been observed in the recent LHCb measurement \cite{Aaij:2017tyk} where the deviation from the SM value is $1.7\sigma$. These deviations, though statistically small and unable to provide unambiguous signal of New Physics (NP) can not be completely ignored. It is rather natural to explore all possible avenues of flavor physics to draw a definite conclusion on the existence of NP.

In the past several decades, the main phenomenological focus in this context has been on exclusive and inclusive decays of $B$ meson. The recent $B$ physics program at LHCb involves in addition to precise measurements of $B$ decay modes, study of baryonic decays such as $b\to s\ell^+\ell^-$ mediated $\Lambda_b\to\Lambda\ell^+\ell^-$. The important difference between $\Lambda_b$ baryon and the $B$ meson decay modes is that due to the spin of the $\Lambda_b$ baryon, its decay distributions are generally more involved which makes the baryonic modes phenomenologically rich. At present, data on baryonic modes are rather limited and only recently LHCb has reported a measurement of several $\Lambda_b\to\Lambda\mu^+\mu^-$ observables where the branching ratio is found to be lower than the SM prediction at low $q^2$ and higher at large $q^2$ \cite{Aaij:2015xza}. Earlier, $\Lambda_b \to \Lambda\mu^+\mu^-$ branching ratio was measured by CDF \cite{Aaltonen:2011qs}.

In the SM, angular distributions of $\Lambda_b\to\Lambda\ell^+\ell^-$ have been studied in \cite{Gutsche:2013pp} and  \cite{Boer:2014kda}. Other studies in the SM and in the context of NP have been reported in Refs.~\cite{Aslam:2008hp,Wang:2008sm,Huang:1998ek,Chen:2001ki,Chen:2001zc,Aliev:2010uy,Roy:2017dum,Sahoo:2016nvx}. The $\Lambda_b\to\Lambda\mu^+\mu^-$ angular distributions for polarized $\Lambda_b$ were recently worked out in \cite{Blake:2017une}. In Ref.~\cite{Meinel:2016grj} the LHCb data on $\Lambda_b\to\Lambda\mu^+\mu^-$ decay in combination with $b\to s\mu^+\mu^-$ data from $B$ meson decays were used to perform a model-independent fit to Wilson coefficients. One of the challenging aspect of $\Lambda_b\to\Lambda\ell^+\ell^-$ decay is to estimate the hadronic $\Lambda_b\to \Lambda$ form factors. Recent progress in this respect includes the lattice QCD results at low and intermediate recoil \cite{Detmold:2016pkz,Detmold:2012vy} and the QCD sum-rules analysis of spectator-scattering corrections to form factor relations which are valid at large recoil \cite{Feldmann:2011xf}. Also, better theoretical understanding of light-cone distribution amplitudes of $\Lambda_b$ baryon has recently been achieved in \cite{Ali:2012pn, Bell:2013tfa, Braun:2014npa}. In the large recoil region, the form factors have been calculated in the light cone sum-rules (LCSR) \cite{Wang:2015ndk,Wang:2008sm}. 

In this paper, we supplement the previous studies with a model independent analysis of $\Lambda_b\to\Lambda\mu^+\mu^-$ for the most general effective Hamiltonian involving new vector and axial vector (VA), scalar and pseudo-scalar (SP), and tensor (T) operators for unpolarized $\Lambda_b$ and massless leptons. We work in the helicity basis and derive the expressions of helicity amplitudes for this Hamiltonian. We present expressions of the double differential branching ratio with respect to dilepton invariant mass squared $q^2$ and cosine of lepton angle $\theta_\ell$, and then define forward-backward asymmetry involving leptons and longitudinal polarization fraction. We use helicity parametrization of $\Lambda_b\to\Lambda$ hadronic matrix elements. At large recoil, which correspond to low $q^2$, the form factors are taken from recent calculations in the light cone sum-rules \cite{Wang:2015ndk} and at low recoil, which correspond to large $q^2$, we use results from the calculations in lattice QCD \cite{Detmold:2016pkz}.

The paper is organized as follows. In Sec.~\ref{subsec:effHam} we write down the most general effective Hamiltonian involving VA, SP and T operators. After describing the decay kinematics in Sec.~\ref{sec:kin}, we describe the helicity formalism in Sec.~\ref{sec:helAmp}. The form factors are discussed in Sec.~\ref{sec:ff} and the model independent analysis is performed in Sec.~\ref{sec:analysis}. The results are summarized in Sec.~\ref{sec:sum}. The paper is supplemented with a number of appendixes which contain the details of the calculations.

\section{Effective Hamiltonian \label{subsec:effHam}}
The most general low energy effective Hamiltonian for rare $|\Delta B|=|\Delta S|=1$ transition involving VA, SP and T operators is
\begin{equation}\label{eq:Heff}
\mathcal{H} = \mathcal{H}^{\rm VA} + \mathcal{H}^{\rm SP} + \mathcal{H}^{\rm T}\, ,
\end{equation}
where $\mathcal{H}^{\rm VA}, \mathcal{H}^{\rm SP}$ and $\mathcal{H}^{\rm T}$ are
\begin{eqnarray}
\mathcal{H}^{\rm VA} &=& -\frac{4G_F}{\sqrt{2}}V_{tb}V_{ts}^\ast \frac{\alpha_e}{4\pi} \Big[\mC_9^{\rm eff} \bar{s} \gamma^\mu P_Lb \bar{\ell}\gamma_\mu\ell + \mC_{10} \bar{s}\gamma^\mu P_L b \bar{\ell}\gamma_\mu \gamma_5 \ell - \frac{2m_b}{q^2}\mC_7^{\rm eff} \bar{s}iq_\nu\sigma^{\mu\nu}P_R b \bar{\ell}\gamma_\mu\ell \, \nn\\ 
&+&  \mC_V \bar{s} \gamma^\mu P_L b \bar{\ell}\gamma_\mu\ell + \mC_A \bar{s}\gamma^\mu P_L b \bar{\ell}\gamma_\mu \gamma_5 \ell + \mC^\prime_V \bar{s} \gamma^\mu P_R b \bar{\ell}\gamma_\mu\ell + \mC^\prime_A \bar{s}\gamma^\mu P_R b \bar{\ell}\gamma_\mu \gamma_5 \ell\Big]\, ,\\
\mathcal{H}^{\rm SP} &=& -\frac{4G_F}{\sqrt{2}}V_{tb}V_{ts}^\ast \frac{\alpha_e}{4\pi} \Big[ \mC_S^\prime \bar{s} P_L b \bar{\ell}\ell + \mC_P^\prime \bar{s} P_L b \bar{\ell}\gamma_5 \ell + \mC_S \bar{s} P_R b \bar{\ell}\ell + \mC_P \bar{s} P_R b \bar{\ell}\gamma_5 \ell\Big]\, ,\\
\mathcal{H}^{\rm T} &=& -\frac{4G_F}{\sqrt{2}}V_{tb}V_{ts}^\ast \frac{\alpha_e}{4\pi} \Big[ \mC_T \bar{s} \sigma^{\mu\nu} b \bar{\ell} \sigma_{\mu\nu} \ell +  \mC_{T5} \bar{s} \sigma^{\mu\nu} b \bar{\ell}\sigma_{\mu\nu}\gamma_5\ell    \Big]\, .
\end{eqnarray}
Here $G_F$ is the Fermi-constant, $\alpha_e$ is the fine structure constant, $V_{tb}V_{ts}^\ast$ are the Cabibbo-Kobayashi-Maskawa(CKM) elements, $P_{L,R}=(1\mp\gamma_5)/2$ are the chiral projection operators, and $\sigma_{\mu\nu}=i[\gamma_\mu,\gamma_\nu]$/2. The $b$-quark mass multiplying the dipole operator is in the $\overline{\text{MS}}$-mass scheme. In the SM, $\mC_V=\mC_V^\prime$ $=\mC_A=\mC_A^\prime=$ $\mC_S=\mC_S^\prime=\mC_P=$ $\mC_P^\prime=\mC_T=\mC_{T5}=0$. The expressions of $\mC_{9}^{\rm eff}$ and other SM Wilson coefficients are given in Appendix \ref{app:C910}. In this article, we neglect the mass of the leptons. 

In the Hamiltonian (\ref{eq:Heff}), we assume only the factorizable quark-loop contributions to four-quark operators and gluonic operator which are conveniently absorbed in the Wilson coefficients $\mC^{\rm eff}_{7,9}$. For simplicity we ignore the non-factorizable corrections which are expected to play a non-negligible role, particularly at low $q^2$ \cite{Beneke:2001at,Beneke:2004dp}.

\section{Kinematics of $\Lambda_b\to\Lambda\ell^+\ell^-$ decay \label{sec:kin}}
The momenta and spins variables for the different particles in the decay process are assigned as follows
\begin{equation}
\Lambda_b(p,s_p ) \to \Lambda(k,s_k) \ell^+(q_1) \ell^-(q_2)\, .
\end{equation}
Here $p,k,q_1$ and $q_2$ are the momenta of the parent baryon $\Lambda_b$, daughter baryon $\Lambda$ and the negatively and positively charged leptons, respectively, and $s_{p,k}$ are the projections of the baryonic spins on to the $z$-axis in their respective rest frames. 

For future convenience, we define the four momentum of the dilepton pair as
\begin{equation}
q^\mu = q_1^\mu + q_2^\mu\, .
\end{equation}
We assume that in the rest frame of $\Lambda_b$, the daughter baryon $\Lambda$ travels along the positive $z$-direction and the lepton pair travels in the negative $z$-direction. So that in $\Lambda_b$ rest frame ($\Lambda_b{\rm RF}$) we have 
\begin{equation}
q^\mu|_{\Lambda_b{\rm RF}} = (q^0, 0, 0, -|q|)\, ,\quad k^\mu|_{\Lambda_b{\rm RF}} =((m_{\Lambda_b}-q^0), 0, 0, +|q|)\, ,
\end{equation}
where 
\begin{equation}
q^0\Big|_{\Lambda_b{\rm RF}} = \frac{\mmLb-\mmL+q^2}{2\mLb}\, ,\quad |q|\Big|_{\Lambda_b{\rm RF}} = \frac{\sqrt{\lambda(\mmLb,\mmL,q^2)}}{2\mLb}\, ,
\end{equation}
and we have defined $\lambda(a,b,c)=a^2+b^2+c^2-2(ab+bc+ca)$.
The $\Lambda_b\to\Lambda(\to N\pi)\ell^+\ell^-$ decay can be described in terms of four independent kinematic variables which we choose as the dilepton invariant mass squared $q^2$ and three angles $\theta_\ell$, $\theta_\Lambda$ and $\phi$. The angles are defined as follows: $\theta_\ell$ is the angle made by the negatively charged lepton $\ell^-$ with the $+z$-direction in the dilepton rest frame, $\theta_\Lambda$ is made by $N$ with respect to the $+z$-direction in the $N\pi$ rest frame, and $\phi$ is the angle between the $\ell^+\ell^-$ and $N\pi$ decay planes.

\section{The Helicity Formalism \label{sec:helAmp}}
\subsection{Decay process \label{sec:hel}}
We derive the decay distribution in the helicity formalism where we restrict ourselves to unpolarized $\Lambda_b$ since the polarization has been measured by LHCb to be very small \cite{Aaij:2013oxa}. In the rest frame of the $\Lambda_b$, the two-fold differential decay distribution in terms of $q^2$ and the angle $\theta_\ell$ can be represented as
\begin{equation}
\frac{d\Gamma}{dq^2 d\cos\theta_\ell} = \frac{1}{2m^3_{\Lambda_b}} \frac{2\sqrt{\lambda(\mmLb,\mmL,q^2)}}{(8\pi)^3}\frac{1}{2s_p+1}\sum_{\lambda_1,\lambda_2}\sum_{s_p,s_k}\big|\mathcal{M}^{\lambda_1,\lambda_2}(s_p,s_k)\big|^2\, ,
\end{equation}
where $\lambda_{1,2}$ are the helicities of the final state leptons. The $\Lambda_b \to \Lambda\ell^+\ell^-$ helicity amplitudes $\mathcal{M}^{\lambda_1,\lambda_2}(s_p,s_k)$ corresponding to the effective Hamiltonian (\ref{eq:Heff}) consist of contributions from VA, SP and T operators
\begin{equation}
\mathcal{M}^{\lambda_1,\lambda_2}(s_p,s_k) = \mathcal{M}^{\lambda_1,\lambda_2}_{\rm VA}(s_p,s_k) + \mathcal{M}^{\lambda_1,\lambda_2}_{\rm SP}(s_p,s_k) + \mathcal{M}^{\lambda_1,\lambda_2}_{\rm T}(s_p,s_k)\, ,
\end{equation}
which are defined as
\begin{align}
& \mathcal{M}^{\lambda_1,\lambda_2}_{\rm VA}(s_p,s_k) = -\frac{G_F}{\sqrt{2}} V_{tb}V_{ts}^\ast \frac{\alpha_e}{4\pi} \sum_\lambda \eta_\lambda \bigg[ H^{L,s_p,s_k}_{\rm VA,\lambda} L^{\lambda_1,\lambda_2}_{L,\lambda}\ + H^{R,s_p,s_k}_{\rm VA,\lambda} L^{\lambda_1,\lambda_2}_{R,\lambda} \bigg]\, ,\\
& \mathcal{M}^{\lambda_1,\lambda_2}_{\rm SP}(s_p,s_k) = -\frac{G_F}{\sqrt{2}} V_{tb}V_{ts}^\ast \frac{\alpha_e}{4\pi} \bigg[ H_{\rm SP}^{L,s_p,s_k} L^{\lambda_1,\lambda_2}_L + H_{\rm SP}^{R,s_p,s_k} L^{\lambda_1,\lambda_2}_R \bigg]\, ,\\
& \mathcal{M}^{\lambda_1,\lambda_2}_{\rm T}(s_p,s_k) = -\frac{2G_F}{\sqrt{2}} V_{tb}V_{ts}^\ast \frac{\alpha_e}{4\pi} \sum_{\lambda,\lambda^\prime} \eta_\lambda \eta_{\lambda^\prime} \bigg[ H^{L,s_p,s_k}_{{\rm T},\lambda \lambda^\prime} L^{\lambda_1,\lambda_2}_{L,\lambda \lambda^\prime} + H^{R,s_p,s_k}_{{\rm T},\lambda \lambda^\prime} L^{\lambda_1,\lambda_2}_{R,\lambda \lambda^\prime} \bigg]\, .
\end{align}
Here $\lambda,\lambda^\prime$ are the polarization states of the virtual gauge boson that decays in to dilepton pair, $\eta_t = 1$ and $\eta_{\pm1,0} = -1$. $H^{L(R)}$ and $L_{L(R)}$ are the hadronic and leptonic helicity amplitudes corresponding the left- $(L)$ and right-handed $(R)$ chiralities of the lepton currents. 

The hadronic helicity amplitudes are the projections of $\Lambda_b\to\Lambda$ matrix elements on the direction of the polarization of virtual gauge boson. For VA, SP and T operators these are 
\begin{eqnarray}\label{eq:Hl}
H^{L(R),s_p,s_k}_{{\rm VA},\lambda} &=& \bar{\epsilon}^\ast_\mu(\lambda) \big{\langle}\Lambda(k,s_k)\big| \bigg[ \Big( (\mC_9^{\rm eff} \mp \mC_{10}) + (\mC_V \mp \mC_A) \Big)\bar{s} \gamma^\mu (1-\gamma_5)b \, \nn\\ &+& (\mC_V^\prime \mp \mC_A^\prime) \bar{s}\gamma^\mu(1+\gamma_5)b - \frac{2m_b}{q^2} \mC_7^{\rm eff} \bar{s}iq_\nu\sigma^{\mu\nu} (1+\gamma_5) b \bigg] \big|\Lambda_b(p,s_p)\big\rangle \, ,\\
H^{L(R),s_p,s_k}_{\rm SP} &=& \big\langle \Lambda(k,s_k) \big| \bigg[ (\mC_S^\prime \mp \mC_P^\prime) \bar{s}(1-\gamma_5)b \,\nn\\&+& (\mC_S \mp \mC_P) \bar{s}(1+\gamma_5)b \bigg] \big| \Lambda_b(p,s_p) \big\rangle\, ,\\
\label{eq:HTll}
H^{L(R),s_p,s_k}_{{\rm T},\lambda\lambda^\prime} &=& i\bar{\epsilon}^\ast_\mu(\lambda) \bar{\epsilon}^\ast_\nu(\lambda^\prime) \big\langle \Lambda(k,s_k) \big| \bar{s}\sigma^{\mu\nu} b \big| \Lambda_b(p,s_p) \big\rangle (\mC_T \mp \mC_{T5}) \, .
\end{eqnarray}
Here $\bar{\epsilon}_\mu(\lambda^{(\prime)})$ denote the polarization vectors of the virtual gauge boson. Our choice for the polarization vectors are summarized in Appendix \ref{app:pol}. Similarly, the leptonic helicity amplitudes are 
\begin{align}\label{eq:Ldef1}
& L^{\lambda_1,\lambda_2}_{L(R)} = \langle \bar{\ell}(\lambda_1)\ell(\lambda_2) | \bar{\ell} (1\mp\gamma_5) \ell | 0\rangle\, , \\
& L^{\lambda_1,\lambda_2}_{L(R),\lambda} = \bar{\epsilon}^\mu(\lambda) \langle \bar{\ell}(\lambda_1) \ell(\lambda_2) | \bar{\ell} \gamma_\mu (1\mp\gamma_5) \ell | 0\rangle\, ,\\
\label{eq:Ldef2}
&L^{\lambda_1,\lambda_2}_{L(R),\lambda \lambda^\prime} = -i\bar{\epsilon}^\mu(\lambda) \bar{\epsilon}^\nu(\lambda^\prime) \langle \bar{\ell}(\lambda_1) \ell(\lambda_2) | \bar{\ell} \sigma_{\mu\nu} (1\mp\gamma_5) \ell | 0\rangle\, .
\end{align}
The tensor amplitudes are anti-symmetric under the exchange of $\lambda$ and $\lambda^\prime$, \emph{i.e.,} $	L^{\lambda_1,\lambda_2}_{L(R),\lambda^\prime \lambda} = - L^{\lambda_1,\lambda_2}_{L(R),\lambda \lambda^\prime}$.

\subsection{Hadronic Helicity amplitudes \label{sec:helhad}}
The $\Lambda_b \to \Lambda$ hadronic matrix elements for different operators are defined in terms of ten helicity form factors $f^V_{t,0,\perp}$, $f^A_{t,0,\perp}$, $f^T_{0,\perp}$, $f^{T5}_{0,\perp}$ \cite{Feldmann:2011xf} and spinor matrix elements. The definitions are summarized in Appendix~\ref{sec:hme} and the spinor matrix elements for different combinations of spin orientations are worked out in Appendix~\ref{app:spinor}. Using these results we write down the expressions of the helicity amplitudes defined in Eqs.~(\ref{eq:Hl})-(\ref{eq:HTll}) for different operators. For VA operators the non-vanishing amplitudes are
\begin{eqnarray}
H^{L(R),\plpl}_{\rm VA,0} &=& f^V_0(\mLb + \mL) \sqrt{\frac{s_-}{q^2}} \mC^{L(R)}_{\rm VA,+} - f^A_0(\mLb - \mL) \sqrt{\frac{s_+}{q^2}} \mC^{L(R)}_{\rm VA,-} \, \nn\\ &+& \frac{2m_b}{q^2} \bigg( f^T_0 \sqrt{q^2 s_-} - f^{T5}_0 \sqrt{q^2 s_+} \bigg)  \mC_7^{\rm eff}\, ,\\
H^{L(R),\mimi}_{\rm VA,0} &=& f^V_0(\mLb + \mL) \sqrt{\frac{s_-}{q^2}} \mC^{L(R)}_{\rm VA,+} + f^A_0(\mLb - \mL) \sqrt{\frac{s_+}{q^2}} \mC^{L(R)}_{\rm VA,-} \, \nn\\ &+& \frac{2m_b}{q^2} \bigg( f^T_0 \sqrt{q^2 s_-} + f^{T5}_0 \sqrt{q^2 s_+} \bigg)  \mC_7^{\rm eff}\, ,\\
H^{L(R),\mipl}_{\rm VA,+} &=& -f^V_\perp \sqrt{2s_-} \mC^{L(R)}_{\rm VA,+} + f^A_\perp \sqrt{2s_+} \mC^{L(R)}_{\rm VA,-} \, \nn\\ &-& \frac{2m_b}{q^2} \bigg( f^T_\perp(\mLb + \mL) \sqrt{2s_-} - f^{T5}_\perp (\mLb - \mL) \sqrt{2s_+}   \bigg)\mC_7^{\rm eff}\,,~~~\\
H^{L(R),\plmi}_{\rm VA,-} &=& -f^V_\perp \sqrt{2s_-} \mC^{L(R)}_{\rm VA,+} - f^A_\perp \sqrt{2s_+} \mC^{L(R)}_{\rm VA,-} \, \nn\\ &-& \frac{2m_b}{q^2} \bigg( f^T_\perp(\mLb + \mL) \sqrt{2s_-} + f^{T5}_\perp (\mLb - \mL) \sqrt{2s_+}   \bigg)\mC_7^{\rm eff}\,,~~~
\end{eqnarray}
where the variables $s_\pm$ are
\begin{equation}\label{eq:spm}
s_\pm = (\mLb \pm \mL)^2 - q^2\, ,
\end{equation}
and we have defined
\begin{align}
& \mC_{\rm VA,+}^{L(R)} = (\mC_9^{\rm eff}\mp \mC_{10})+(\mC_V\mp \mC_A)+(\mC_V^\prime \mp \mC_A^\prime)\, ,\\
& \mC_{\rm VA,-}^{L(R)} = (\mC_9^{\rm eff}\mp \mC_{10})+(\mC_V\mp \mC_A)-(\mC_V^\prime \mp \mC_A^\prime)\, .
\end{align}
The non-vanishing helicity amplitudes for the SP operators are
\begin{eqnarray}
H^{L(R),\plpl}_{\rm SP} &=& \bigg( f^A_t \frac{\mLb + \mL}{m_b}\sqrt{s_-} + f^V_t \frac{\mLb - \mL}{m_b} \sqrt{s_+} \bigg)(\mC_S^\prime \mp \mC_P^\prime)\, \nn\\ &+& \bigg( -f^A_t \frac{\mLb + \mL}{m_b} \sqrt{s_-} + f^V_t \frac{\mLb - \mL}{m_b} \sqrt{s_+} \bigg) (\mC_S \mp \mC_P)\,,~~\\
H^{L(R),\mimi}_{\rm SP} &=& \bigg( -f^A_t \frac{\mLb + \mL}{m_b}\sqrt{s_-} + f^V_t \frac{\mLb - \mL}{m_b} \sqrt{s_+} \bigg)(\mC_S^\prime \mp \mC_P^\prime)\, \nn\\ &+& \bigg( f^A_t \frac{\mLb + \mL}{m_b} \sqrt{s_-} + f^V_t \frac{\mLb - \mL}{m_b} \sqrt{s_+} \bigg) (\mC_S \mp \mC_P)\,,
\end{eqnarray}
where we have neglected the mass of the strange quark. Finally, for the T operators the non-vanishing amplitudes are
\begin{eqnarray}
H^{L(R),\plpl}_{{\rm T},+-} &=& - H^{L(R)\mimi}_{{\rm T},+-} = -f^{T5}_0 \sqrt{s_+} (\mC_T \mp \mC_{T5})\, ,\\
H^{L(R),\mipl}_{{\rm T},+0} &=& H^{L(R)\plmi}_{{\rm T},-0} = f^{T5}_\perp (\mLb - \mL) \sqrt{\frac{2s_+}{q^2}} (\mC_T \mp \mC_{T5})\, ,~~\\
H^{L(R),\mipl}_{{\rm T},+t} &=& H^{L(R)\plmi}_{{\rm T},-t} = f^{T5}_\perp (\mLb + \mL) \sqrt{\frac{2s_-}{q^2}}(\mC_T \mp \mC_{T5})\, ,\\
H^{L(R),\plpl}_{{\rm T},0t} &=& H^{L(R)\mimi}_{{\rm T},0t} = -f^T_0 \sqrt{s_-}(\mC_T \mp \mC_{T5})\, .
\end{eqnarray}
The tensor amplitudes are anti-symmetric under the exchange of $\lambda$ and $\lambda^\prime$, \emph{i.e.,} $H^{L(R)}_{{\rm T},\lambda^\prime\lambda}=-H^{L(R)}_{{\rm T},\lambda\lambda^\prime}$. 


\subsection{Leptonic Helicty amplitudes \label{sec:hellep}} 
Using the representations of the lepton spinors given in Appendix~\ref{sec:llRF}, we calculate the leptonic helicity amplitudes defined in Eqs.~(\ref{eq:Ldef1})-(\ref{eq:Ldef2}). Neglecting the mass of the leptons, the expressions of the non-vanishing amplitudes for the SP and VA operators are
\begin{align}
& L^{\plpl}_L = - L^{\mimi}_R = 2\sqrt{q^2}\, ,\nn\\
& L^{\plmi}_{R,+} = -L^{\mipl}_{L,-} = - \sqrt{2q^2} (1 - \cos\theta_\ell)\, ,\\
%
& L^{\mipl}_{L,+} = -L^{\plmi}_{R,-} = + \sqrt{2q^2} (1 + \cos\theta_\ell)\, ,\quad L^{\mipl}_{L,0} = L^{\plmi}_{R,0} = 2\sqrt{q^2} \sin\theta_\ell\, ,\nn
\end{align}
and for the tensor operators the expressions are
\begin{align}
& L^{\plpl}_{L,+-} = L^{\mimi}_{R,+-} = -L^{\plpl}_{L,0t} = L^{\mimi}_{R,0t} = - 2 \sqrt{q^2} \cos\theta_\ell\, , \nn\\
& L^{\plpl}_{L,+0} = L^{\mimi}_{R,+0} = L^{\plpl}_{L,-0} = L^{\mimi}_{R,-0} = + \sqrt{2q^2} \sin\theta_\ell\, ,\\
& L^{\plpl}_{L,+t} = -L^{\mimi}_{R,+t} = -L^{\plpl}_{L,-t} = L^{\mimi}_{R,-t} = - \sqrt{2q^2} \sin\theta_\ell\, .\nn
%
\end{align}
Other non-vanishing amplitudes can be obtained with the following relation $L^{\lambda_1,\lambda_2}_{L(R),\lambda^\prime\lambda} = - L^{\lambda_1,\lambda_2}_{L(R),\lambda\lambda^\prime}\, .$

\subsection{Differential Distributions and Observables}
With the expressions of the helicity amplitudes in Secs.~\ref{sec:helhad} and \ref{sec:hellep} we write down the two-fold differential branching ratio
\begin{align}\label{eq:2fold}
 \frac{d^2\mathcal{B}}{dq^2 d\cos\theta_\ell} &= N^2(q^2) \Bigg( A^{\rm VA} + A^{\rm SP}  + A^{\rm T} + A^{\rm inter}  \Bigg)\, ,
\end{align} 
where $A^{\rm VA, SP, T}$ correspond to the contributions from VA, SP and T operators respectively and $A^{\rm inter}$ is the interference terms between SP and T contributions. The normalization constant $N(q^2)$ is given by
\begin{equation}
N(q^2) = G_F V_{tb}V_{ts}^\ast \alpha_e  \sqrt{\tau_{\Lambda_b} \frac{q^2\sqrt{\lambda(\mmLb,\mmL,q^2)}}{2^{15} m^3_{\Lambda_b} \pi^5 }  }\, .
\end{equation}
In terms of the hadronic helicity amplitudes the expressions of $A^{\rm VA,SP,T}$ are 
\begin{align}
A^{\rm VA} &= 4 (1 - \ccl) \bigg( \big|H_{\rm VA,0}^{L,\plpl}\big|^2 + \big|H_{\rm VA,0}^{L,\mimi}\big|^2 + \big|H_{\rm VA,0}^{R,\plpl}\big|^2 + \big|H_{\rm VA,0}^{R,\mimi}\big|^2\bigg)\, \nn\\ &  + 2(1-\cl)^2\bigg( \big| H_{\rm VA,-}^{L,\plmi} \big|^2 + \big| H_{\rm VA,+}^{R,\mipl} \big|^2\bigg) \,\nn\\&+ 2(1+\cl)^2 \bigg( \big| H_{\rm VA,-}^{R,\plmi} \big|^2 + \big| H_{\rm VA,+}^{L,\mipl} \big|^2\big| \bigg)\, ,\\[10pt]
A^{\rm SP} & = 4 \bigg( \big| H_{\rm SP}^{L,\plpl} \big|^2 + \big| H_{\rm SP}^{L,\mimi} \big|^2 + \big| H_{\rm SP}^{R,\plpl} \big|^2 + \big| H_{\rm SP}^{R,\mimi} \big|^2\bigg)\,,\\[10pt]
A^{\rm T} & = 8\bigg[ \bigg( \big|H_{\rm T,-0}^{L,\plmi}\big|^2 + \big|H_{\rm T,-t}^{L,\plmi}\big|^2 + \big|H_{\rm T,+0}^{L,\mipl}\big|^2 + \big|H_{\rm T,+t}^{L,\mipl}\big|^2 + \{ L \leftrightarrow R \} \bigg) \,\nn\\ & + \ccl \bigg( 2 \big| H_{\rm T,0t}^{L,\mimi} \big|^2  + 2 \big| H_{\rm T,0t}^{L,\plpl} \big|^2  - \big| H_{\rm T,-0}^{L,\plmi} \big|^2 - \big| H_{\rm T,+0}^{L,\mipl} \big|^2 \nn \\ & - \big| H_{\rm T,-t}^{L,\plmi} \big|^2 - \big| H_{\rm T,+t}^{L,\mipl} \big|^2  + 2 \big| H_{\rm T,+-}^{L,\mimi} \big|^2 + 2 \big| H_{\rm T,+-}^{L,\plpl} \big|^2 + \{ L \leftrightarrow R \}  \bigg) \,\nn\\ & + \ccl \bigg( 4\re\big[H_{\rm T,+-}^{L,\mimi} (H_{\rm T,0t}^{L,\mimi})^\ast\big] + 4\re\big[H_{\rm T,+-}^{L,\plpl} (H_{\rm T,0t}^{L,\plpl})^\ast\big]\, \nn\\ & + 2\re\big[H_{\rm T,-t}^{L,\plmi} (H_{\rm T,-0}^{L,\plmi})^\ast\big] - 2 \re\big[H_{\rm T,+t}^{L,\mipl} (H_{\rm T,+0}^{L,\mipl})^\ast\big] - \{ L \leftrightarrow R \}  \bigg) \,\nn\\ & -  \bigg( 2\re\big[H_{\rm T,-t}^{L,\plmi} (H_{\rm T,-0}^{L,\plmi})^\ast\big] -  2\re\big[H_{\rm T,+t}^{L,\mipl} (H_{\rm T,+0}^{L,\mipl})^\ast\big] -  \{ L \leftrightarrow R \} \bigg) \bigg]\,,
\end{align}
and the interference term is
\begin{align}
A^{\rm inter}_{\rm SP-T} & = -16\cl \Bigg[\bigg( \re\big[H_{\rm T,0t}^{L,\mimi}(H_{\rm SP}^{L,\mimi})^\ast\big] + \re\big[H_{T,0t}^{L,\plpl}(H_{\rm SP}^{L,\plpl})^\ast\big] +\{L \leftrightarrow R \} \bigg) \,\nn\\ &+ \bigg(\re\big[H_{\rm T,+-}^{L,\mimi}(H_{\rm SP}^{L,\mimi})^\ast\big] + \re\big[H_{\rm T,+-}^{L,\plpl}(H_{\rm SP}^{L,\plpl})^\ast\big] - \{L \leftrightarrow R \}  \bigg) \bigg] \, .
\end{align}
As can be seen, there are no interference terms of VA with SP and T contributions. This is because VA-SP and VA-T interferences are proportional to either lepton mass or its squared and therefore vanish in the limit of massless leptons.

By partially integrating the differential distribution (\ref{eq:2fold}) over the angle $\theta_\ell$ various observables can be constructed. Here we recall the definitions of the leptonic forward-backward asymmetry $A^\ell_{\rm FB}$
\begin{equation}
	A^\ell_{\rm FB}(q^2) = \frac{ \displaystyle\int_{-1}^{0} d\cos\theta_\ell \frac{d^2\mathcal{B}}{dq^2d\cos\theta_\ell} - \displaystyle\int_{0}^{+1} d\cos\theta_\ell \frac{d^2\mathcal{B}}{dq^2d\cos\theta_\ell} }{ \displaystyle\int_{-1}^{0} d\cos\theta_\ell \frac{d^2\mathcal{B}}{dq^2d\cos\theta_\ell} + \displaystyle\int_{0}^{+1} d\cos\theta_\ell \frac{d^2\mathcal{B}}{dq^2d\cos\theta_\ell} }\, ,
\end{equation}
and the longitudinal polarization fraction $F_L$
\begin{equation}
	F_L(q^2) = \frac{ \displaystyle\int_{-1}^{+1} d\cos\theta_\ell (2-5\cos^2\theta_\ell) \frac{d^2\mathcal{B}}{dq^2d\cos\theta_\ell} }{ \displaystyle\int_{-1}^{+1} d\cos\theta_\ell \frac{d^2\mathcal{B}}{dq^2d\cos\theta_\ell} }\, .
\end{equation}
In section \ref{sec:analysis} we perform a model-independent analysis of these observables in the presence of the new operators.

\section{$\Lambda_b \to \Lambda$ helicity form factors \label{sec:ff}}
There are ten $q^2$ dependent form factors $f^V_{t,0,\perp}$, $f^A_{t,0,\perp}$, $f^T_{0,\perp}$, $f^{T5}_{0,\perp}$ that parametrize the $\Lambda_b\to\Lambda$ hadronic matrix elements which are summarized in Appendix \ref{sec:hme}. Note that our notations of the form factors is same as in \cite{Boer:2014kda}. Somewhat different notation has been followed in \cite{Detmold:2016pkz} and our notation is related to their's as $f^V_{t,0,\perp} = f_{0,+,\perp}$, $f^A_{t,0,\perp} = g_{0,+,\perp}$, $f^T_{0,\perp}=h_{+,\perp}$ and $f^{T5}_{0,\perp}=\tilde{h}_{+,\perp}$. 

In the large recoil region, we use the LCSR results \cite{Wang:2015ndk} for the form factors. The $q^2$ dependence are obtained from fits of the LCSR results to $z$-parametrization with following form
\begin{equation}
	f^i(q^2) = \frac{f^i(0)}{1-q^2/m^2_{B^\ast_s}} \bigg( 1 + b_1^i [ z(q^2,t_+) - z(0,t_+) ]  \bigg)\, ,
\end{equation}
for the form factors $f^V_0,f^V_\perp, f^T_0$ and $f^T_\perp$,
\begin{equation}
		f^i(q^2) = \frac{f^i(0)}{1-q^2/m^2_{B_s}} \bigg( 1 + b_1^i [ z(q^2,t_+) - z(0,t_+) ]  \bigg)\, ,
\end{equation}
for the form factor $f^A_t$, and
\begin{equation}
f^i(q^2) = f^i(0) \bigg( 1 + b_1^i [ z(q^2,t_+) - z(0,t_+) ]  \bigg)\, ,
\end{equation}
for the form factors $f^V_t, f^A_{0,\perp}, f^{T5}_{0,\perp}$. Here $z(q^2,t_+)$ is defined as
\begin{equation}
	z(q^2,t_+) = \frac{\sqrt{t_+ - q^2} - \sqrt{t_+ - t_0}}{\sqrt{t_+ - q^2} + \sqrt{t_+ - t_0}}\, ,
\end{equation}
where $t_0 = (\mLb - \mL)^2$ and $t_+ = (m_{B_s} + m_\pi)^2$. The values of the form factors at zero momentum transfer $f^i(0)$, the fit parameters $b_1^i$, and the masses of $B_s^{(\ast)}$ are taken from \cite{Wang:2015ndk}. In our numerical analysis at large recoil we restrict ourselves to $ q^2 \leq 7$ GeV$^2$ in dilepton invariant mass squared.

At low recoil region we use lattice QCD results \cite{Detmold:2016pkz} for the form factors. In Ref.~\cite{Detmold:2016pkz} two $z$-parametrization fits, ``nominal" fit and ``higher-order" fit have been used to obtain the $q^2$ dependence. For our numerical analysis, we use their ``higher-order" fit which correspond to the following $z$-parametrization
\begin{equation}
f^i(q^2) = \frac{1}{1-q^2/(m^{f^i}_{\rm pole})^2} \big[ a^{f^i}_0 + a_1^{f^i} z(q^2,t^\prime_+) + a_2^{f^i} (z(q^2,t^\prime_+))^2 \big]\, ,
\end{equation}
where $t^\prime_+ = (m_B + m_K)^2$. The fit parameters are taken from \cite{Detmold:2016pkz}. We note that if both the ``nominal" and ``higher-order" fit are used in the way mentioned in Ref.~\cite{Detmold:2016pkz}, it would lead to different central values and errors in our results.  For numerical analysis, our choice of $q^2$ range for low recoil is $14.18 < q^2 < (\mLb - \mL)^2$ GeV$^2$. 

\begin{figure}[h!]
	\begin{center}
		\includegraphics[scale=0.34]{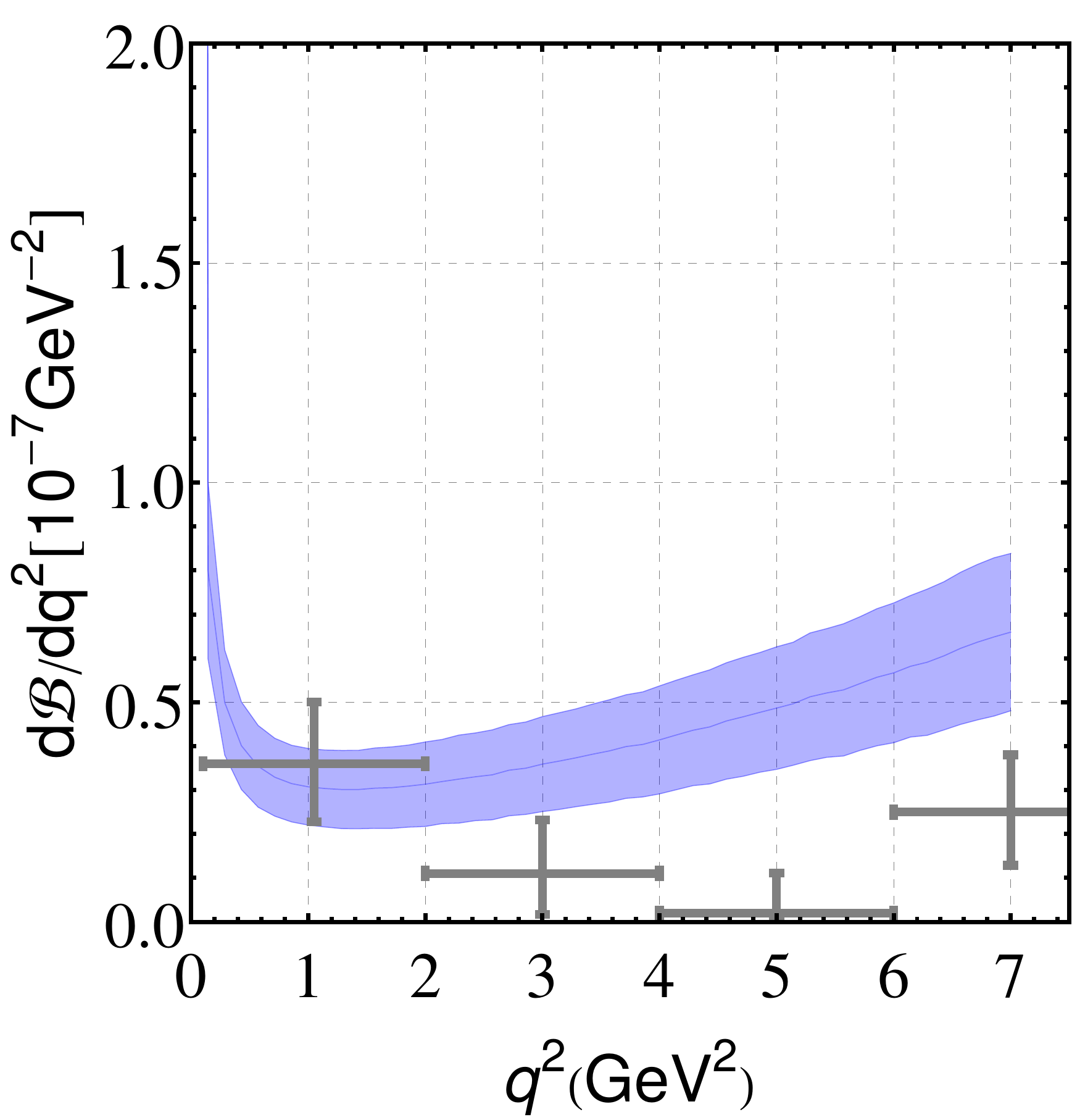}
		\includegraphics[scale=0.34]{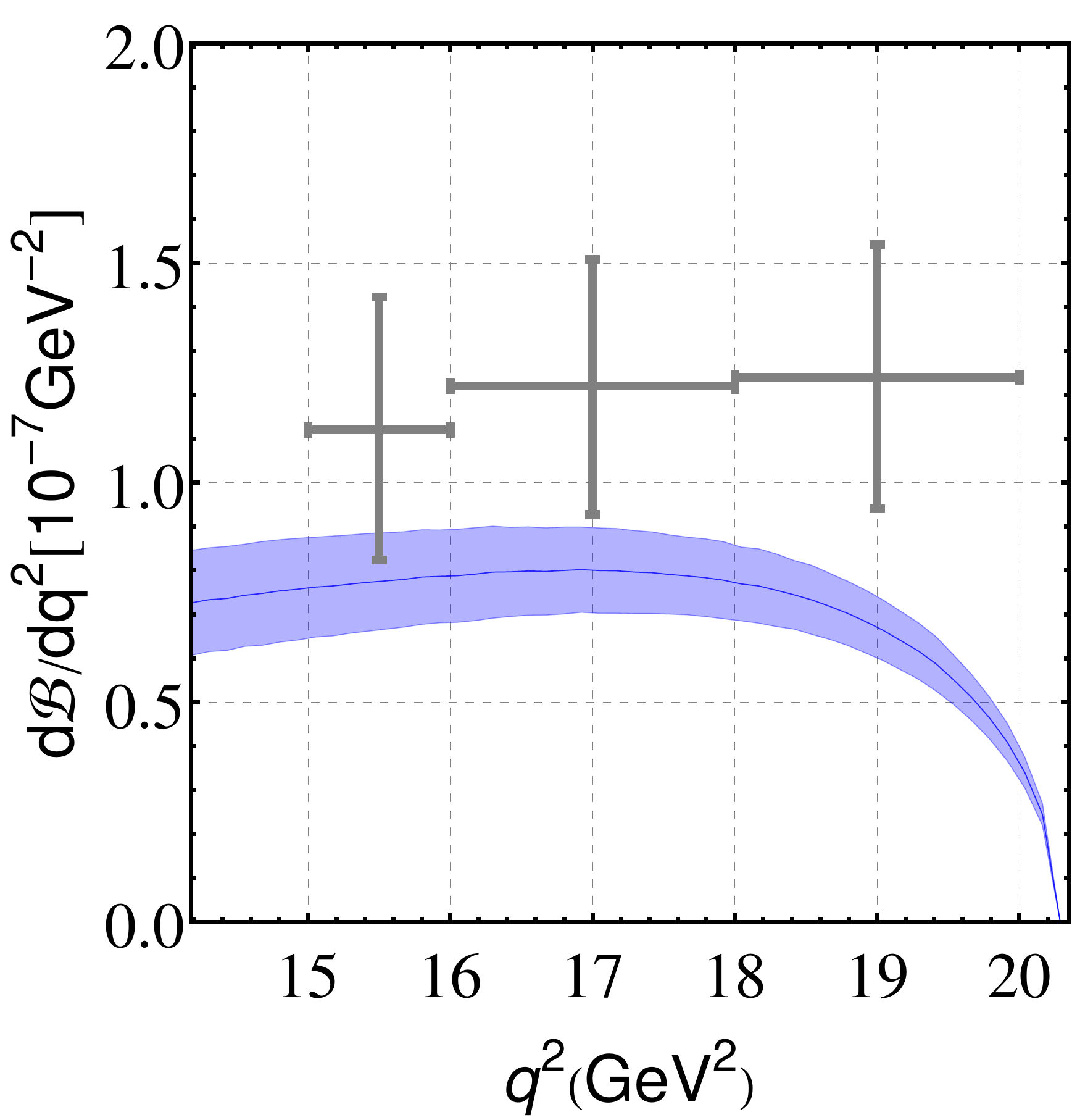}
		\includegraphics[scale=0.35]{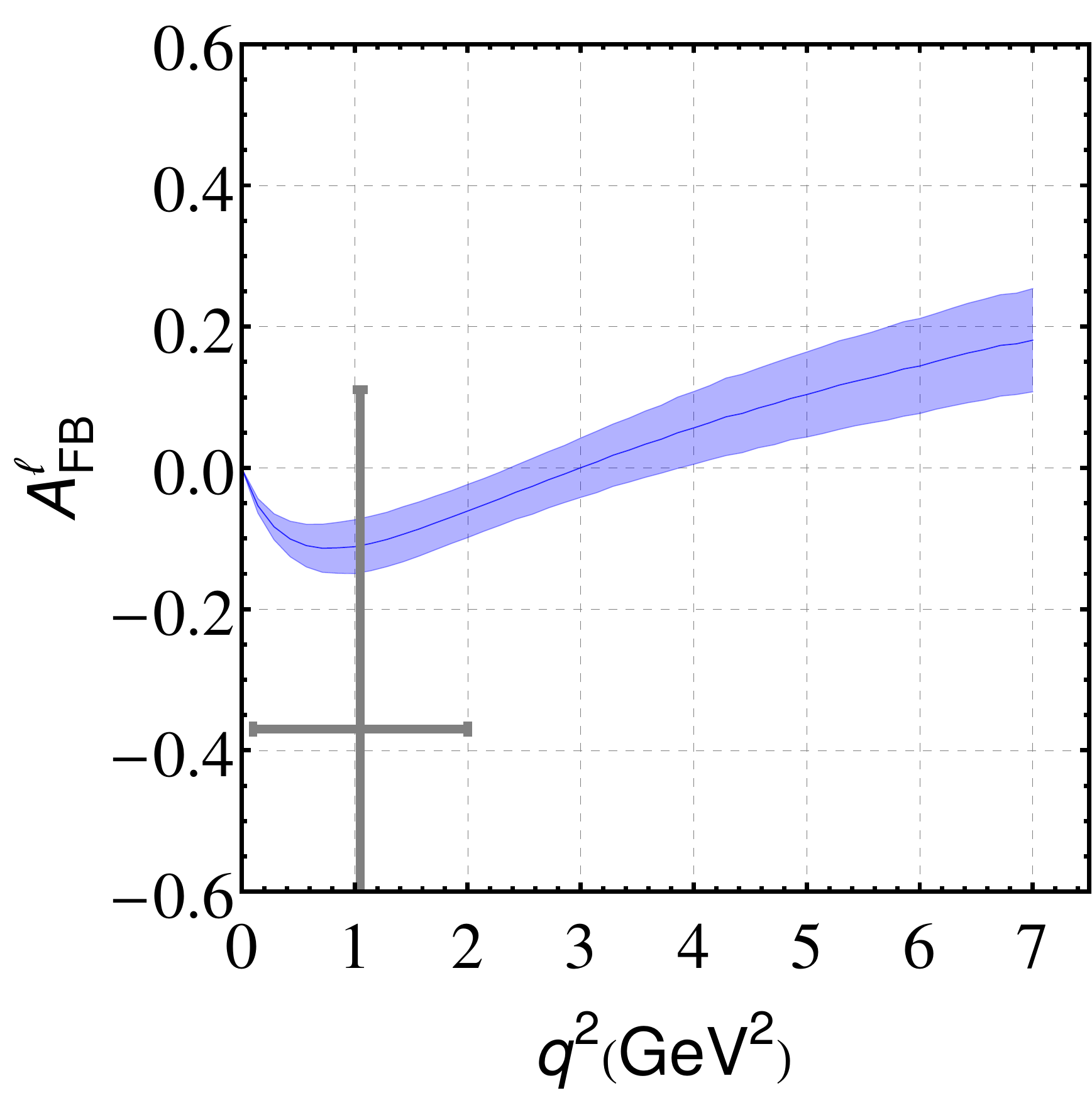}
		\includegraphics[scale=0.35]{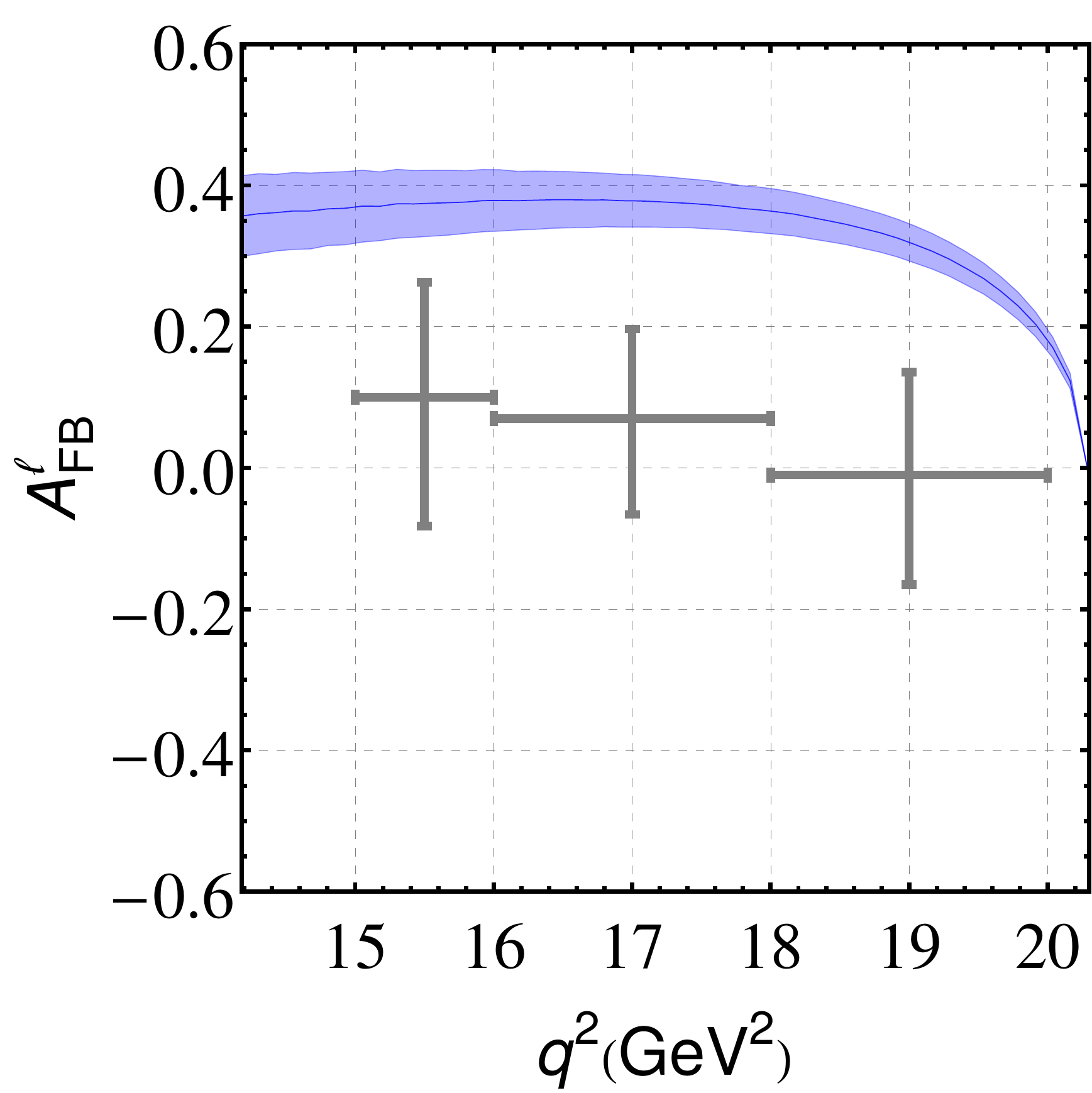}
		\includegraphics[scale=0.34]{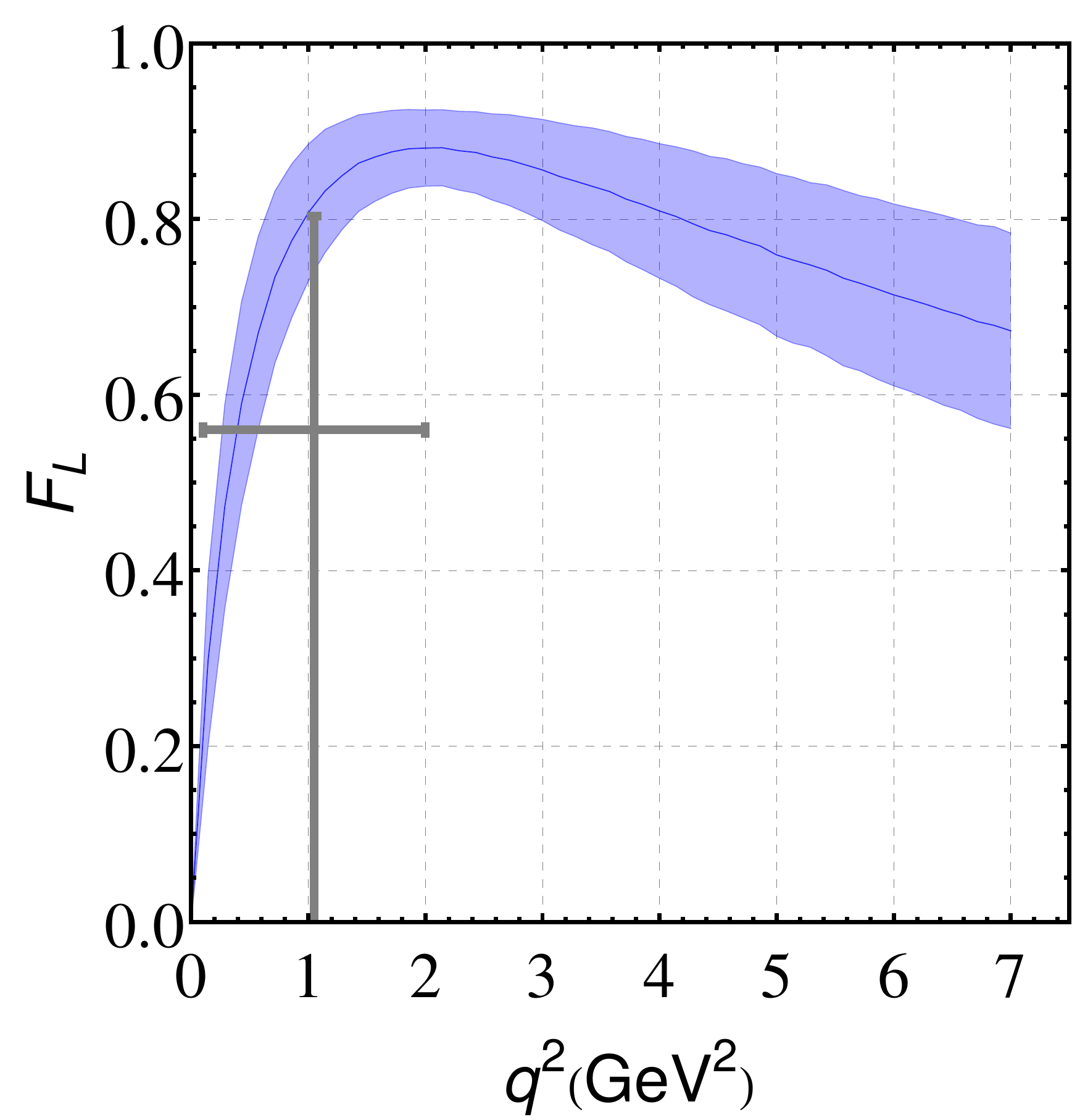}
		\includegraphics[scale=0.34]{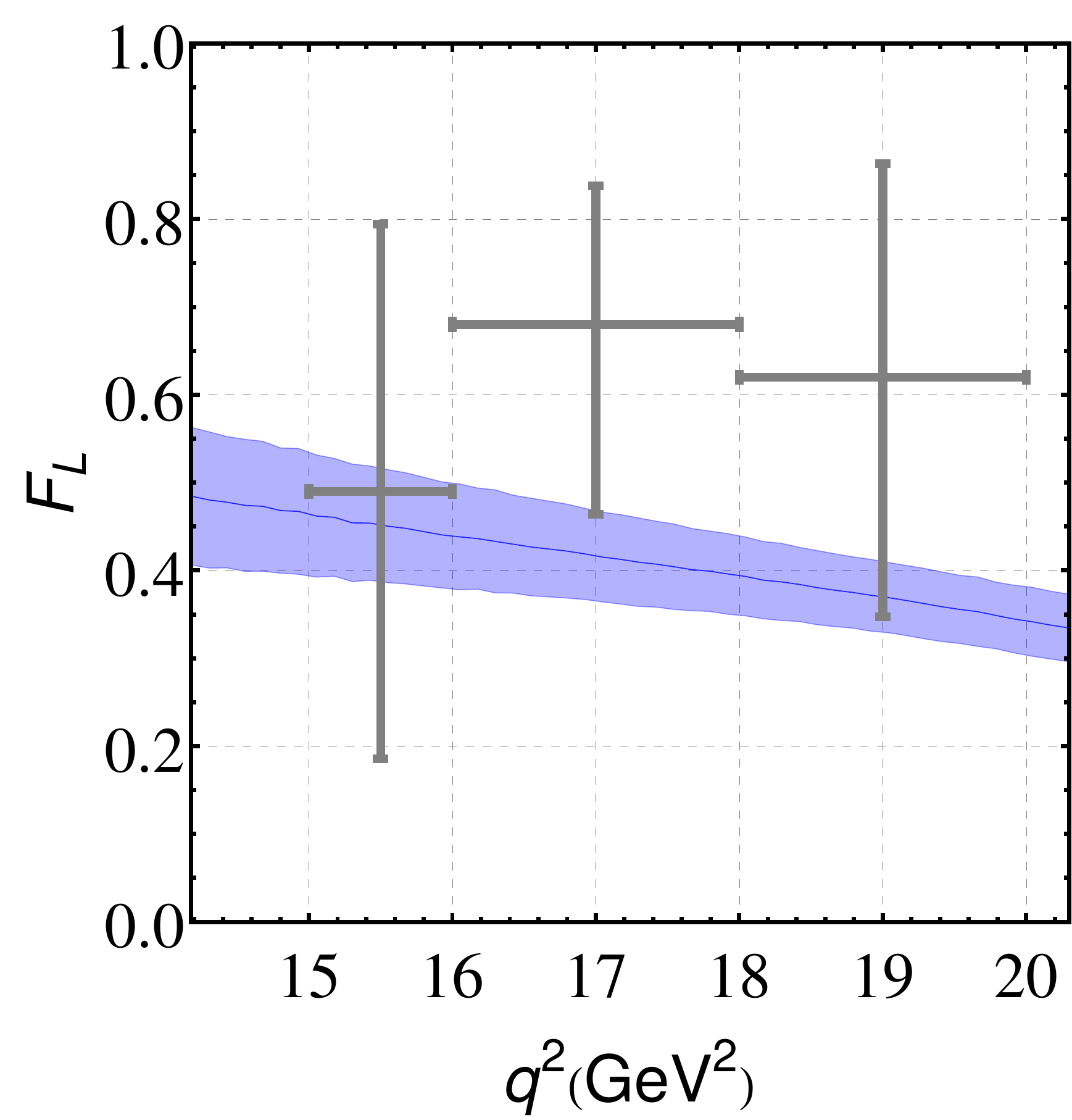}
		\caption{The SM predictions of differential branching ratio, leptonic forward-backward asymmetry $A_{\rm FB}^{\ell}$ and the longitudinal polarization fraction $F_L$. The bands correspond to the uncertainties dominated by the $\Lambda_b \to \Lambda$ form-factors. The points shown by cross marks indicate LHCb data \cite{Aaij:2015xza}.   \label{fig:SM}}
	\end{center}
\end{figure}

With the form factors at hand, we are ready to study the angular observables in the SM. In Fig.~\ref{fig:SM} the differential branching ratio $d\mathcal{B}/dq^2$, leptonic forward-backward asymmetry $A_{\rm FB}^\ell$ and the longitudinal polarization fraction $F_L$ are shown as function of $q^2$. The bands are obtained by varying the form factors and other inputs within their quoted uncertainties. A list of numerical inputs and their values are collected in table~\ref{tab:inputs}.
We have superimposed the latest LHCb data by gray crosses. In our case, the angle $\theta_\ell$ is defined with respect to the negatively charged lepton, where as LHCb choses the positively charged lepton to define the angle $\theta_\ell$. This difference has been taken care of to display the LHCb data on $A^\ell_{\rm FB}$.


\section{Model Independent Analysis \label{sec:analysis}}
In this section we study the impacts of NP couplings on the differential branching ratio, forward-backward asymmetry and longitudinal polarization fraction. We sequentially discuss the consequences of the presence of (i) only $\mC_{V,A}^{(\prime)}$ couplings, (ii) only $\mC_{S,P}^{(\prime)}$ couplings, (iii) only $\mC_{T,T5}$ couplings, and (iv) combinations of $\mC_{V,A}^{(\prime)}$ $\mC_{S,P}^{(\prime)}$ and $\mC_{T,T5}$ couplings. All the NP couplings are considered real. The values of numerical inputs used in our analysis are collected in the table \ref{tab:inputs} in Appendix \ref{sec:inputs}.

\subsection{Only VA couplings are present \label{sec:VA}}
\begin{figure}[h!]
	\begin{center}
		\includegraphics[scale=0.33]{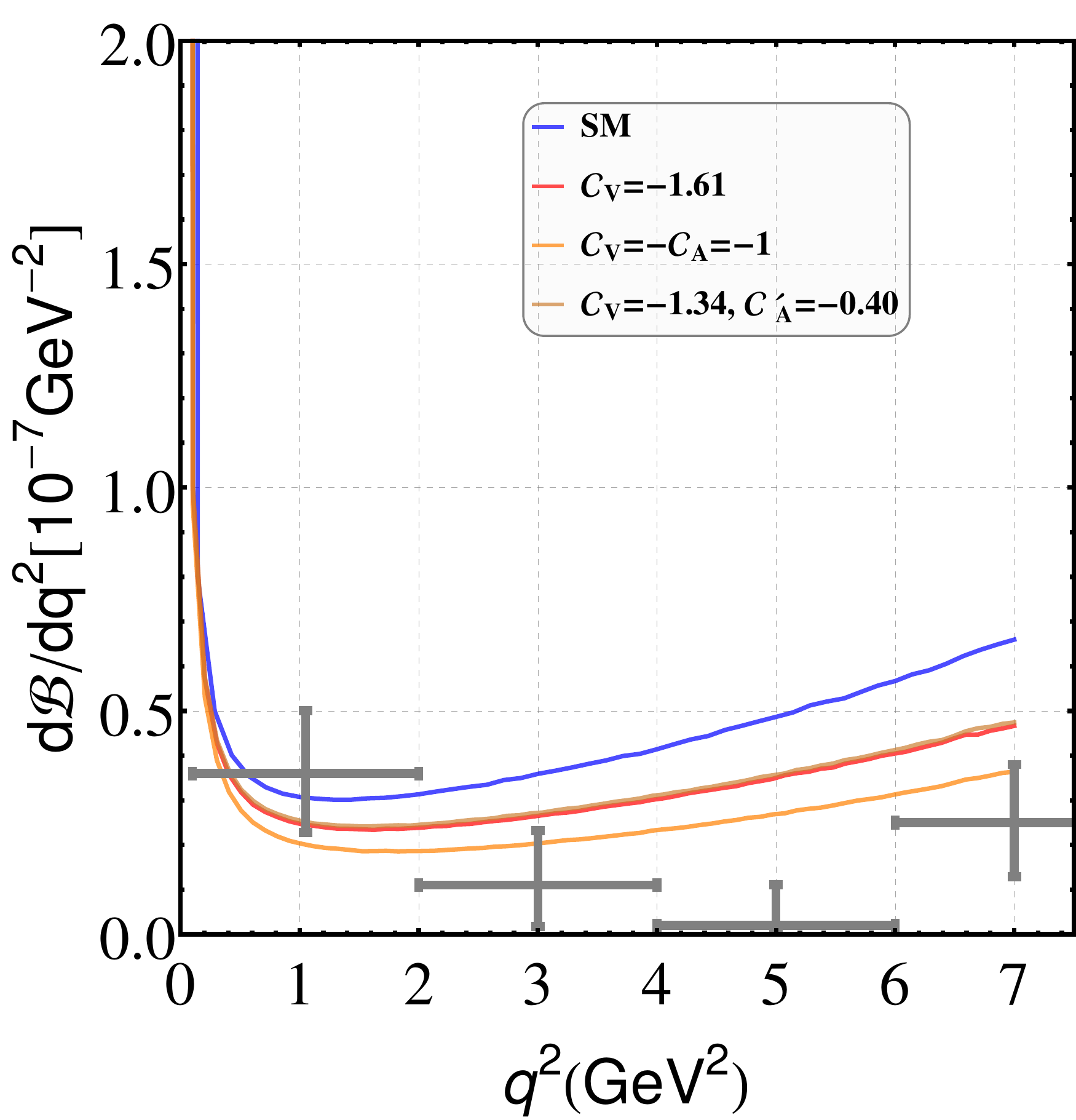}
		\includegraphics[scale=0.33]{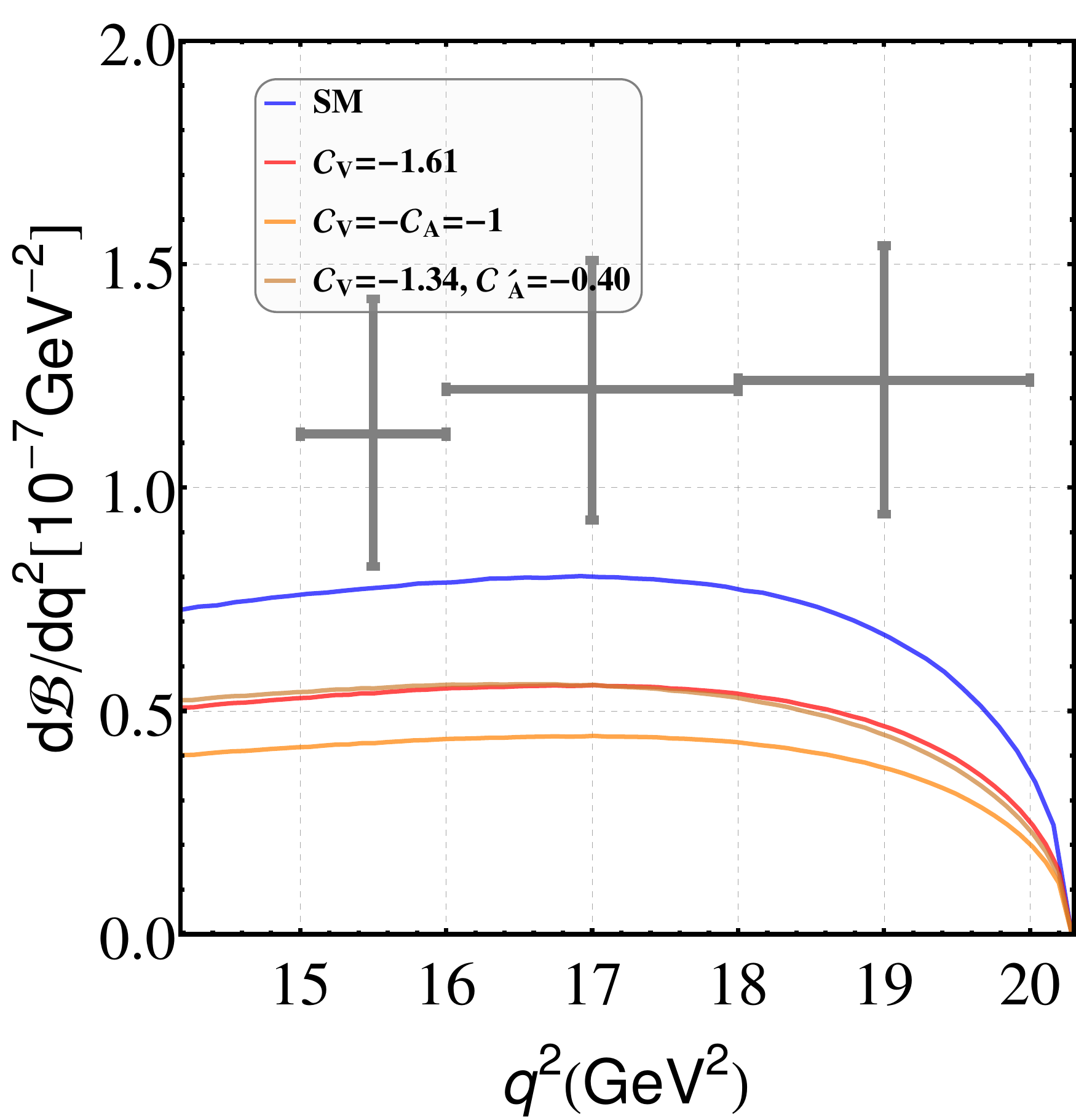}
		\includegraphics[scale=0.35]{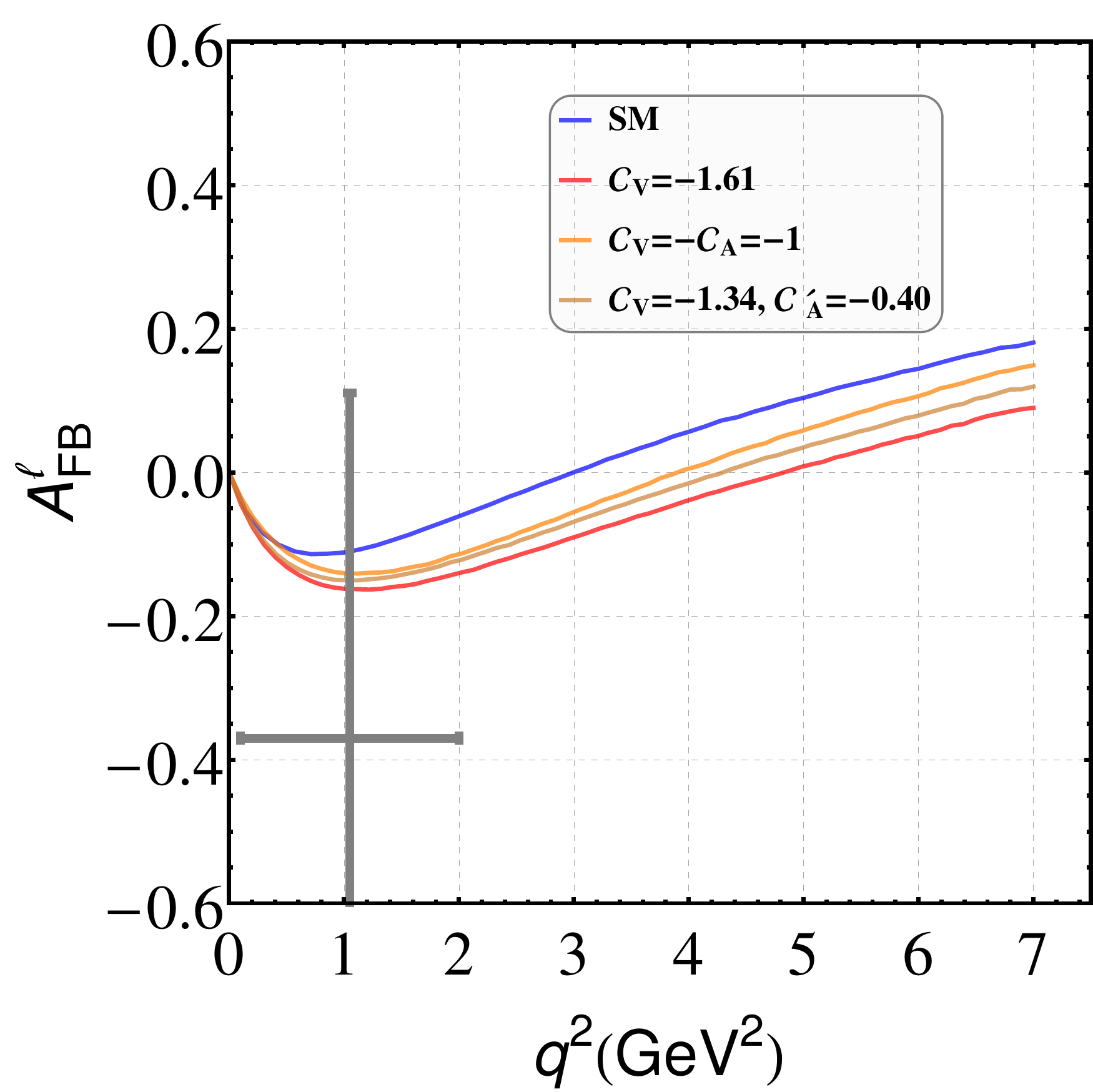}
		\includegraphics[scale=0.33]{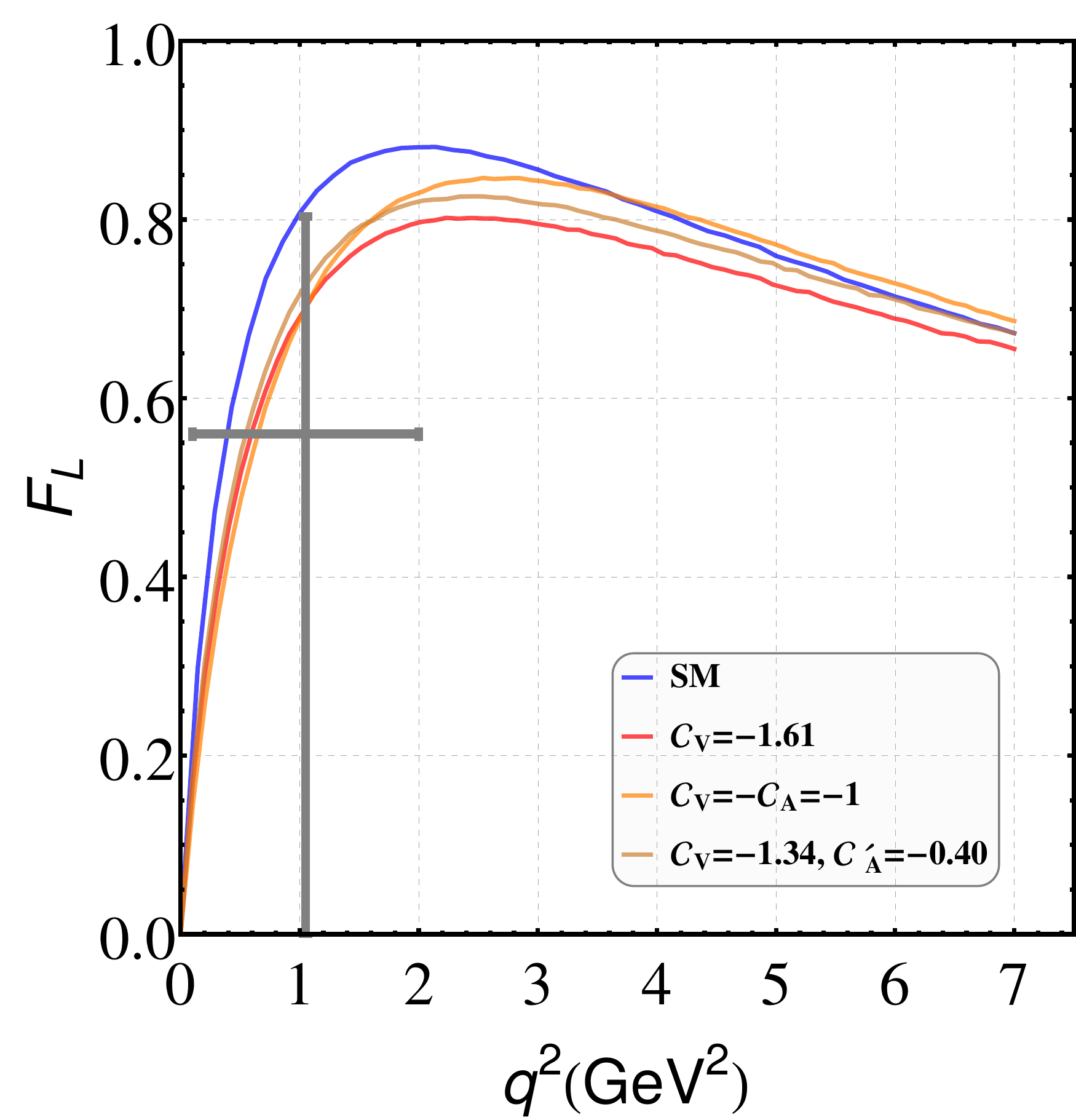}
		\caption{The $\Lambda_b \to \Lambda\mu^+\mu^-$ observables in the SM (blue) and for representative sets of new VA couplings (shades of red). The points shown by cross marks indicate LHCb data \cite{Aaij:2015xza}.  \label{fig:VA}}
	\end{center}
\end{figure}

There exist a wealth of data on $b\to s\mu^+\mu^-$ transitions from various $B$ meson decay modes some of which show deviations from the SM predictions. In the past few years a number of global fits have appeared to interpret the deviations through new VA couplings. If such new couplings are indeed present, it is worth studying the consequences on $\Lambda_b\to\Lambda\mu^+\mu^-$ observables. We do so by first summarizing the important features when only $\mC_{V,A}^{(\prime)}$ are present. In our discussions, we choose those set of couplings that have large 'pull' in the global fits to $b\to s\mu^+\mu^-$ data. In this context we follow the fits presented in Ref.~\cite{Altmannshofer:2017fio} where it is assumed that NP appear only in the muon modes and the electron modes are SM like. In Fig.~\ref{fig:VA} we show plots with representative sets of couplings which are chosen so as to show the interesting variants. For all our analysis, the signs of the couplings are same as is found in global fits\cite{Altmannshofer:2017fio}. Our observations are as follows
\begin{enumerate}
	\item[]\textbullet~ It was realized very early from global fits that a negative value of $\mC_V$ can easily alleviate the tensions in $b\to s\mu^+\mu^-$ data. We find that for a negative value of $\mC_V$ the branching ratio is reduced at all $q^2$. With only positive $\mC_A$, or, with $\mC_V = -\mC_A$ also the branching ratio is reduced. However, for $\mC_V = \mC_A < 0$, the branching ratio increases for all $q^2$. 
	\item[]\textbullet~ For a negative value of $\mC_V$, the zero-crossing of leptonic forward-backward asymmetry $A^\ell_{\rm FB}$ is shifted to the higher values of $q^2$ than in SM. For $\mC_V = \pm\mC_A$ also, the zero-crossing points are shifted in this direction. However, if only $\mC_A$ is present, the zero-crossing point is SM like, which is understood from the fact that in the SM the zero-crossing point is proportional to $\mC_7^{\rm eff}/\mC_9^{\rm eff}$. 
\end{enumerate}
There are other combinations of VA couplings that can be considered, such as only $\mC_{V,A}^\prime$ or $\mC_V^\prime = \pm \mC_A^\prime$. In global fits these cases have very small `pull' value and we therefore do not discuss these further. 

We also examined four pairs of uncorrelated couplings, $(\mC_V,\mC_A)$, $(\mC_V,\mC_V^\prime)$, $(\mC_V,\mC_A^\prime)$ and $(\mC_V^\prime,\mC_A)$ having large pull values in global fits \cite{Altmannshofer:2017fio}. Here we summarize some of our observations
\begin{enumerate}
	\item[]\textbullet~ We find that the branching ratio is reduced for all these cases.
	\item[]\textbullet~ Apart from when the pair $(\mC_V^\prime,\mC_A)$ is present, the $A^\ell_{\rm FB}$ zero-crossing is shifted to the higher value of $q^2$ for all other pairs. For the case when the pair $(\mC_V^\prime,\mC_A)$ is present, the zero-crossing can shift to lower value of $q^2$, but the effect is negligible unless the couplings are large.
\end{enumerate}

\begin{figure}[h!]
	\begin{center}
		\includegraphics[scale=0.33]{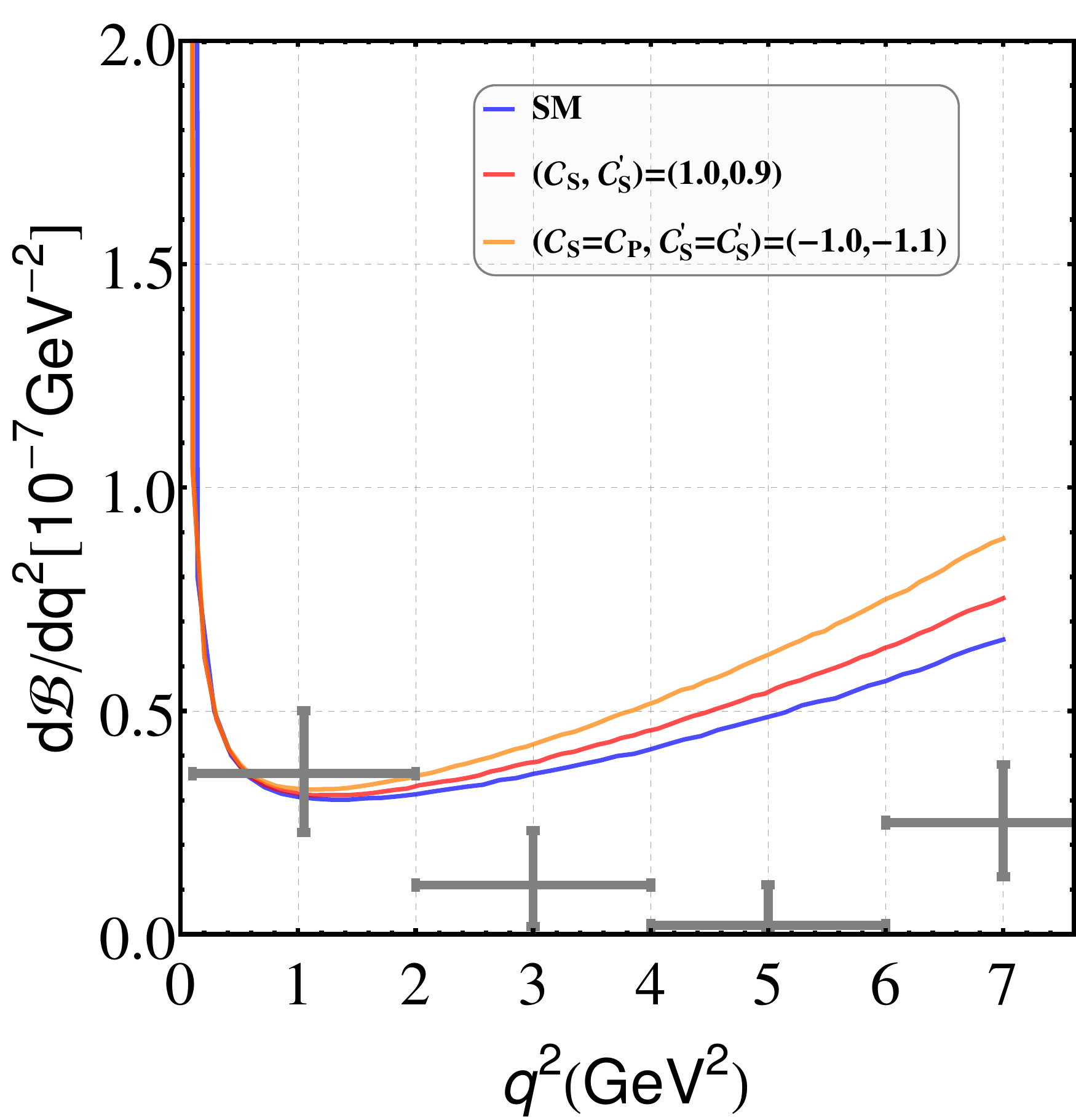}
		\includegraphics[scale=0.33]{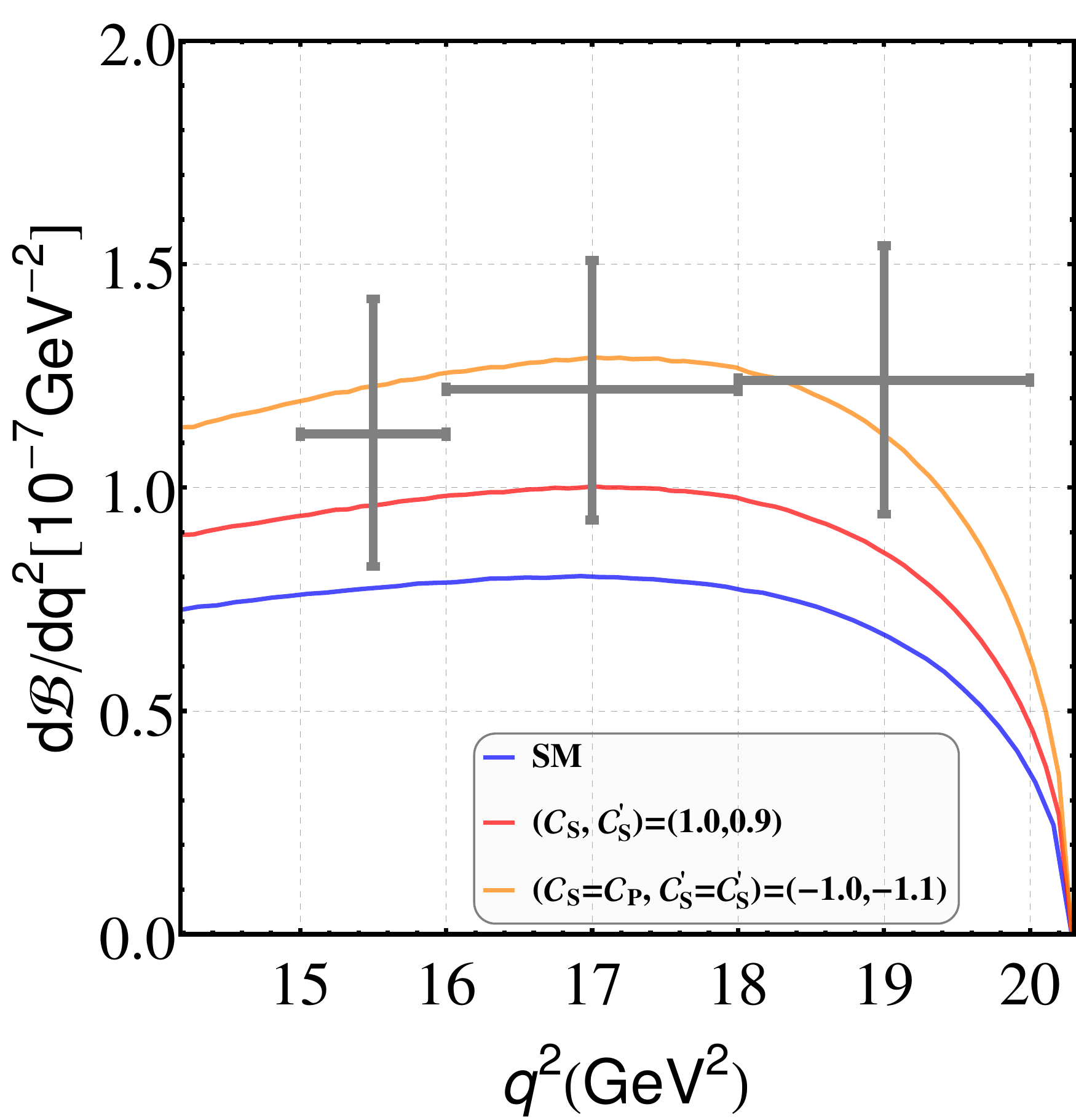}
		\includegraphics[scale=0.34]{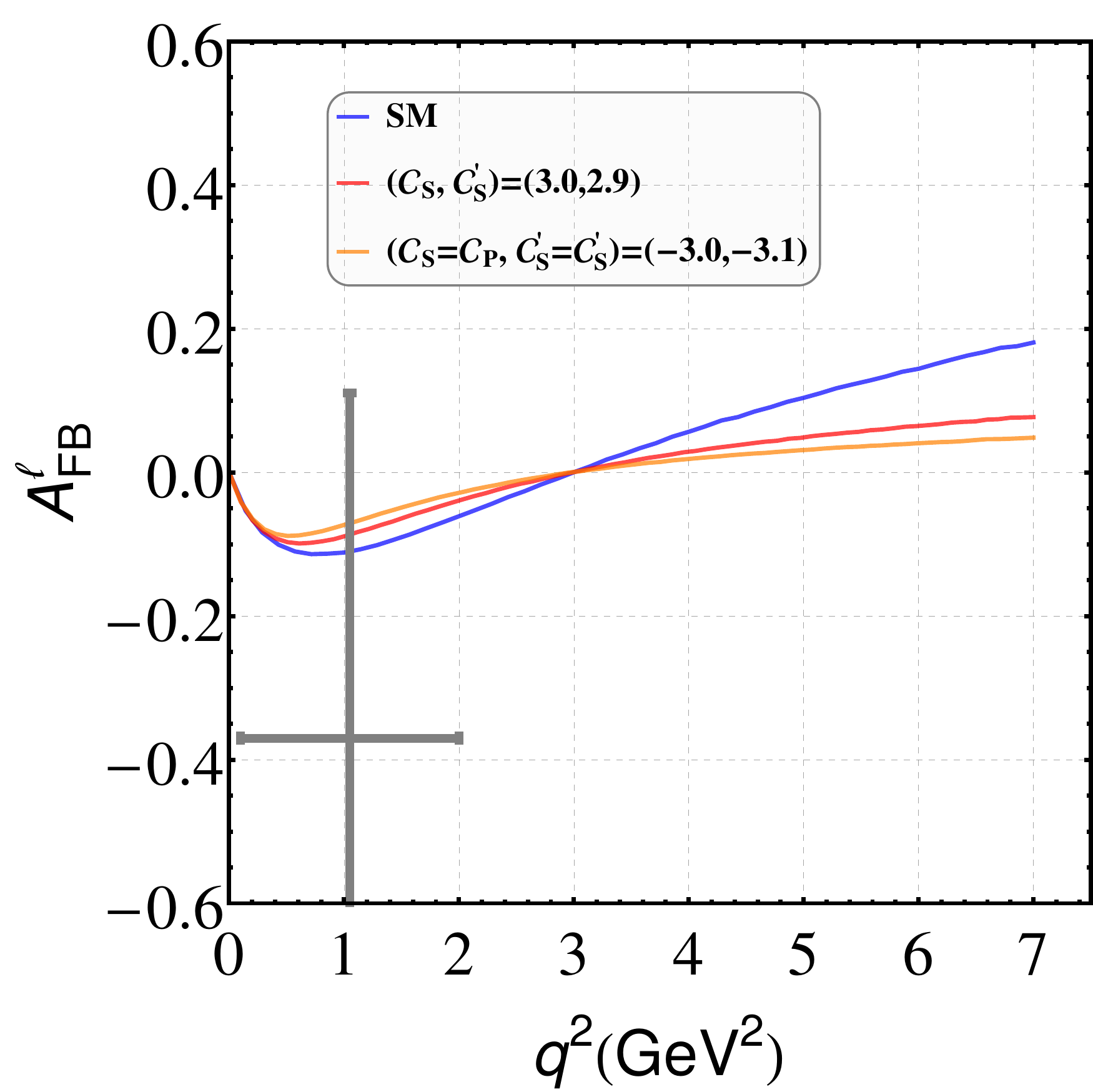}
		\includegraphics[scale=0.34]{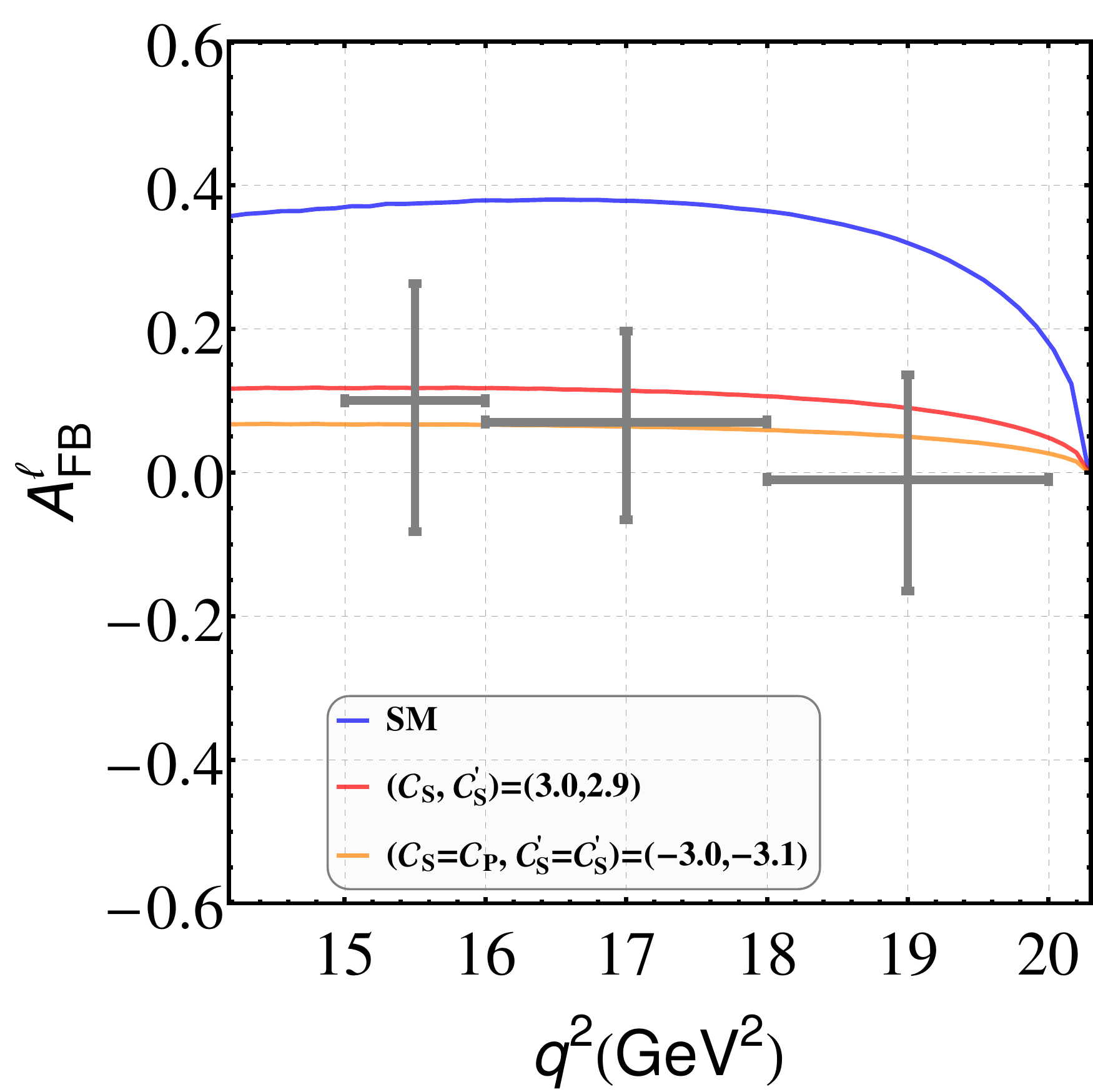}
		\includegraphics[scale=0.33]{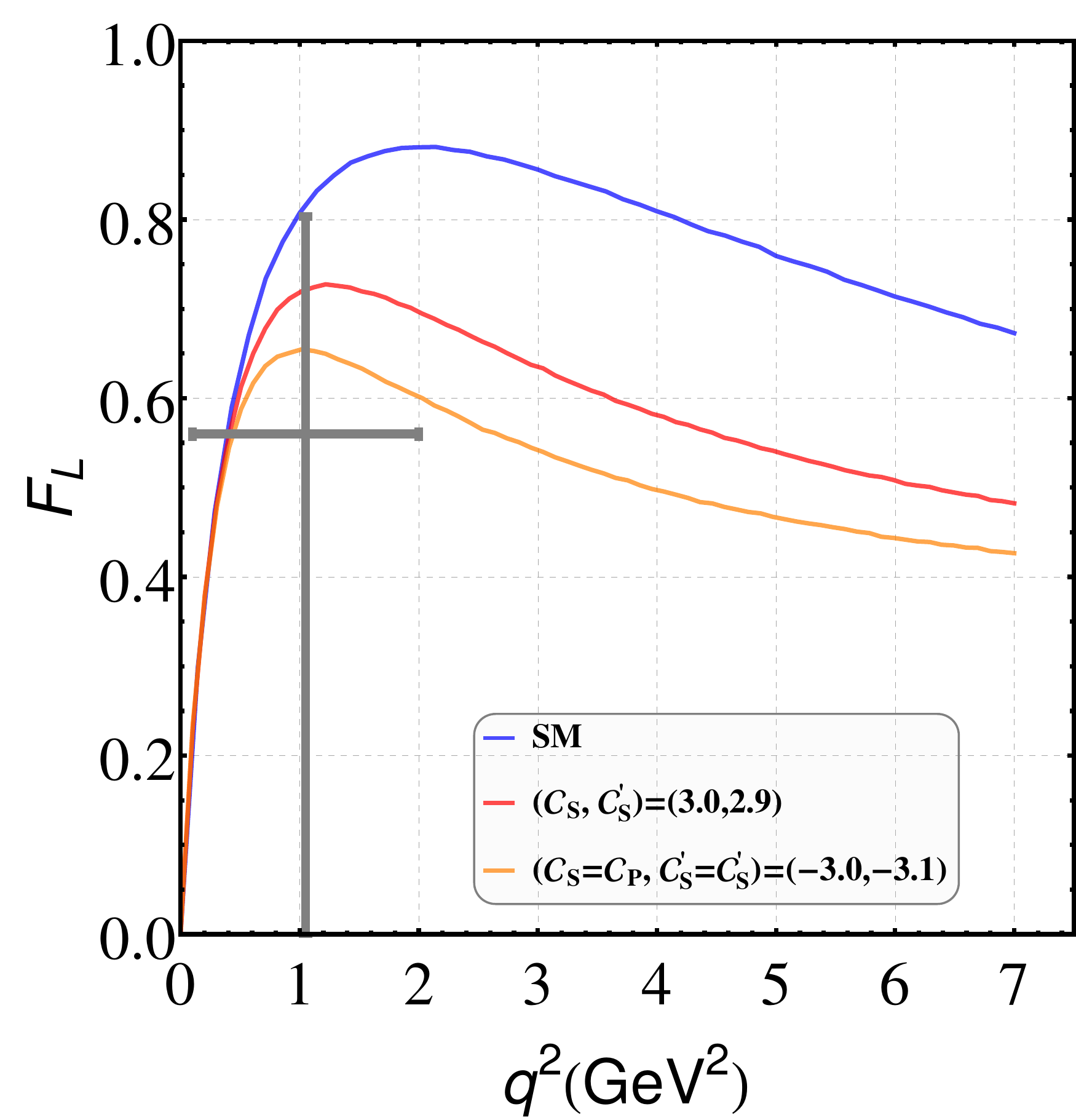}
		\includegraphics[scale=0.33]{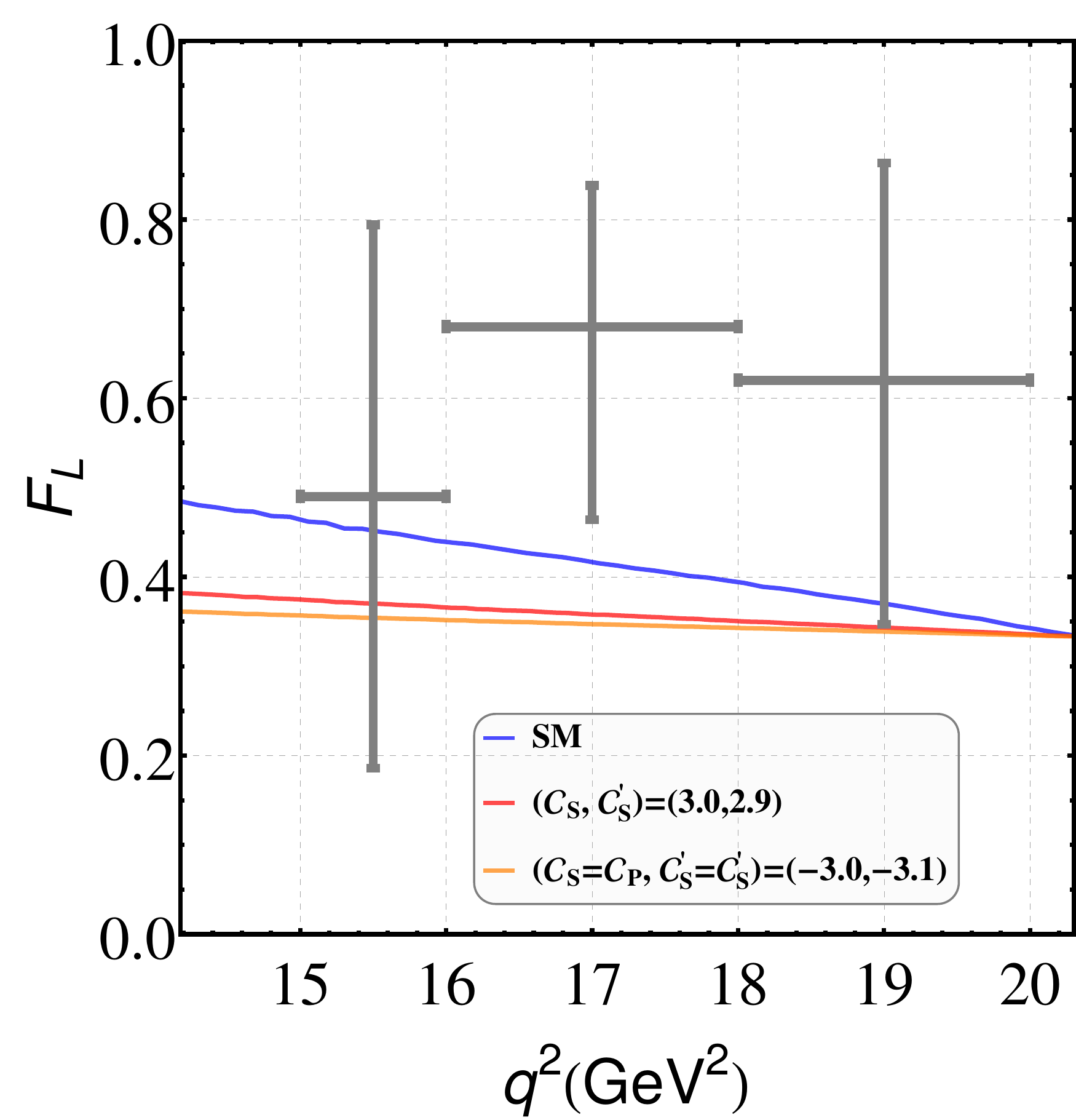}
		\caption{The $\Lambda_b \to \Lambda\mu^+\mu^-$ observables in the SM (blue) and for some new SP couplings (shades of red). The points shown by cross marks indicate LHCb data \cite{Aaij:2015xza}.  \label{fig:SP}}
	\end{center}
\end{figure}

\subsection{Only SP couplings are present}
The scalar operators are constrained from $\bar{B}_s\to \mu^+\mu^-$ branching ratio. To $\bar{B}_s\to\mu^+\mu^-$ decay, there is no contribution from new vector couplings $\mC_V^{(\prime)}$, while the contributions from new axial vector couplings $\mC_A^{(\prime)}$ are suppressed by $m_\mu/m_{B_s}$. There is also no contribution from tensor operators since the matrix element $\langle 0|\bar{s}\sigma^{\mu\nu}b|B_s\rangle$ vanishes. In the presence of SP operators the of $\bar{B}_s \to \mu^+\mu^-$ branching ratio is given by 
\begin{eqnarray}\label{eq:Bsmumu}
	\mathcal{B}(\bar{B}_s\to\mu^+\mu^-) &=& \frac{G_F^2\alpha_e^2 m_{B_s}^5 f_{B_s}^2 \tau_{B_s} }{64\pi^3} |V_{tb}V_{ts}^\ast|^2 \sqrt{1 - \frac{4m_\mu^2}{m_{B_s}^2} } \, \nn\\ &\times& \bigg\{ \bigg( 1 - \frac{4m_\mu^2}{m_{B_s}^2} \bigg) \bigg| \frac{\mC_S - \mC_S^\prime}{m_b} \bigg|^2 + \bigg| \frac{\mC_P - \mC_P^\prime}{m_b} + \frac{2m_\mu}{m_{B_s}^2} \mC_{10}  \bigg|^2 \bigg\}\, .
\end{eqnarray}
 The SM prediction for the branching ratio is $\mathcal{B}^{\rm SM}(\bar{B}_s\to\mu^+\mu^-) = 3.53 \times 10^{-9}$ \cite{Bobeth:2013uxa,Fleischer:2017ltw} while the experimentally measured values by CMS\cite{Chatrchyan:2013bka} and LHCb \cite{Aaij:2017vad} are $3.0^{+1.0}_{-0.9} \times 10^{-9}$, $(3.0 \pm 0.6^{+0.3}_{-0.2}) \times 10^{-9}$, respectively.

As can be seem from Eq.~(\ref{eq:Bsmumu}), the constraints from $\bar{B}_s\to\mu^+\mu^-$ alone can be evaded due to the cancellations between $\mC_{S,P}$ and $\mC_{S,P}^\prime$. The only stringent constraint that can be obtained are $|\mC_S-\mC_S^\prime|\lesssim 0.1$ and $|\mC_P-\mC_P^\prime|\lesssim 0.3$ \cite{Hiller:2014yaa}. Therefore we combine with $\bar{B}_s\to\mu^+\mu^-$ the measurements of $\bar{B}\to X_s\mu^+\mu^-$, which is discussed in the next section, and find that the $\mC_{S,P}^{(\prime)}$ are constrained as
\begin{equation}
	\mC_{S,P}^{(\prime)} \equiv [-4.0,4.0]\, .
\end{equation}
We use these values and our main observations are 
\begin{enumerate}
	\item[]\textbullet~ Branching ratio increases in the presence of SP.
	\item[]\textbullet~ The $A^\ell_{\rm FB}$ zero-crossing always remain SM like. This is understood from the fact that the $A^{\rm SP}$ term does not have any $\cos\theta_\ell$ dependence and therefore do not contribute to the numerator of $A^\ell_{\rm FB}$. This is also the reason why $|A^\ell_{\rm FB}|$ decreases for all $q^2$ as can be seen in Fig.~\ref{fig:SP}.
	\item[]\textbullet~ Longitudinal polarization fraction reduces.
\end{enumerate}

\subsection{Only tensor couplings are present}
\begin{figure}[h!]
	\begin{center}
		\includegraphics[scale=0.33]{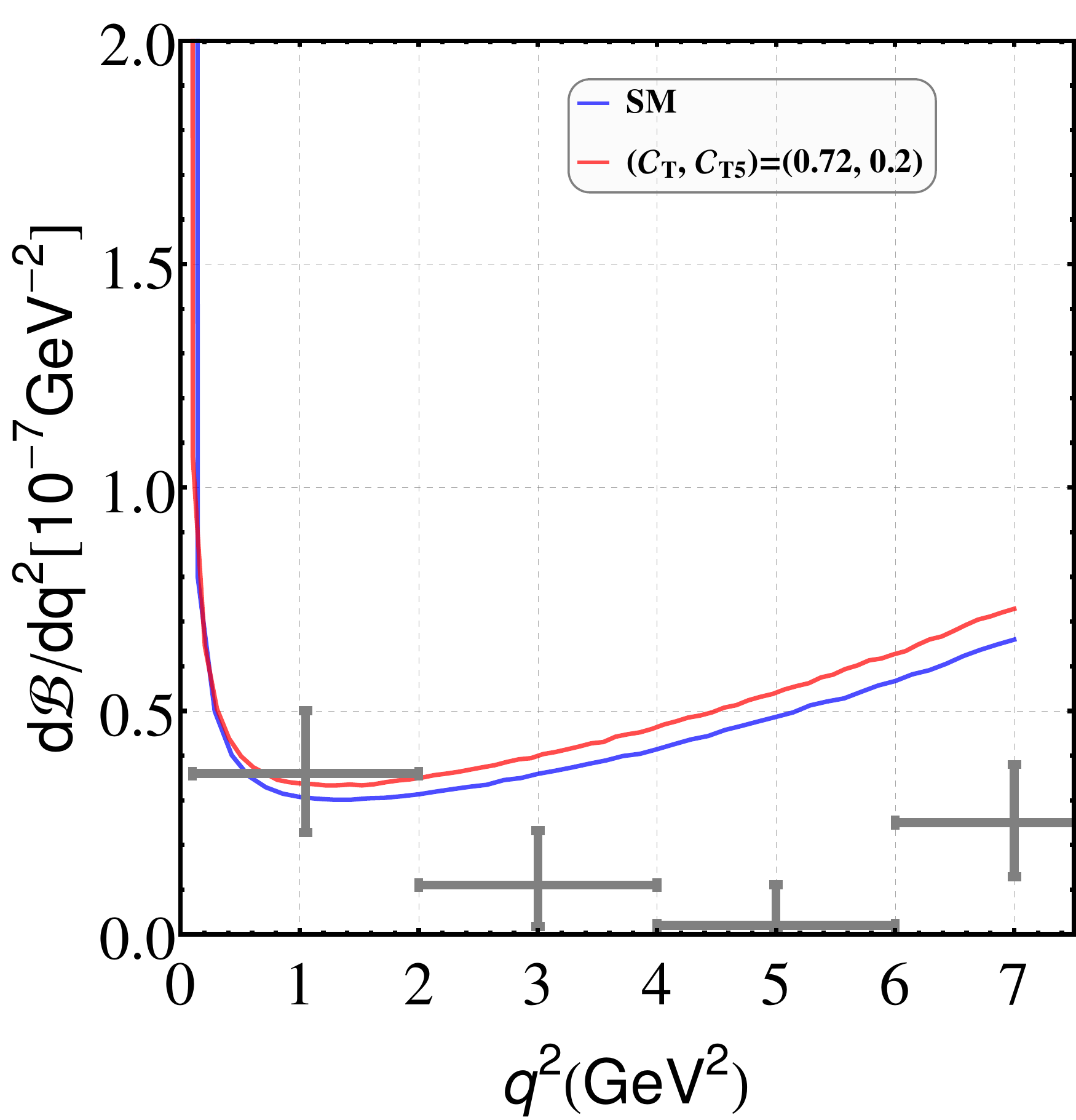}
		\includegraphics[scale=0.33]{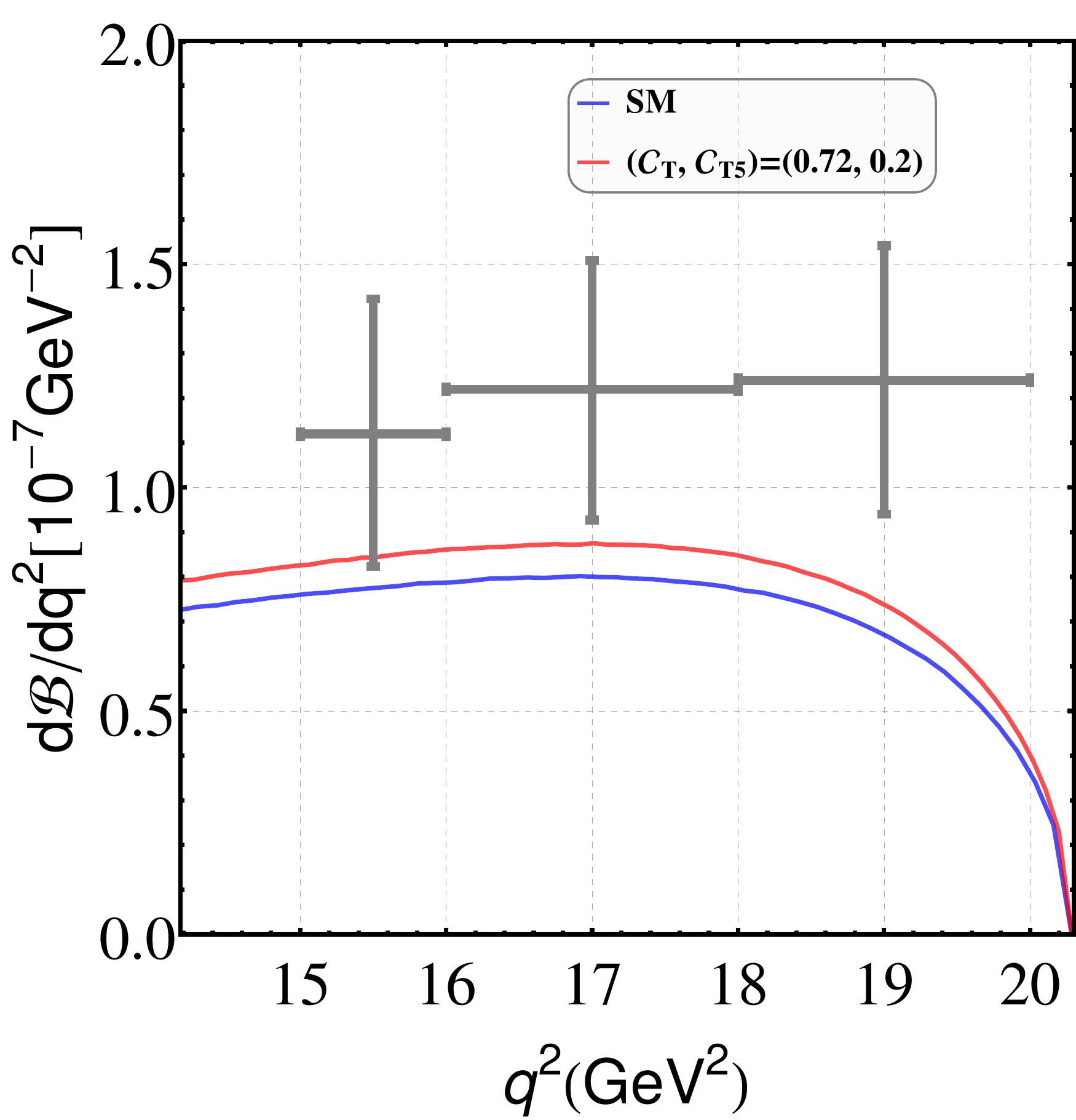}
		\includegraphics[scale=0.33]{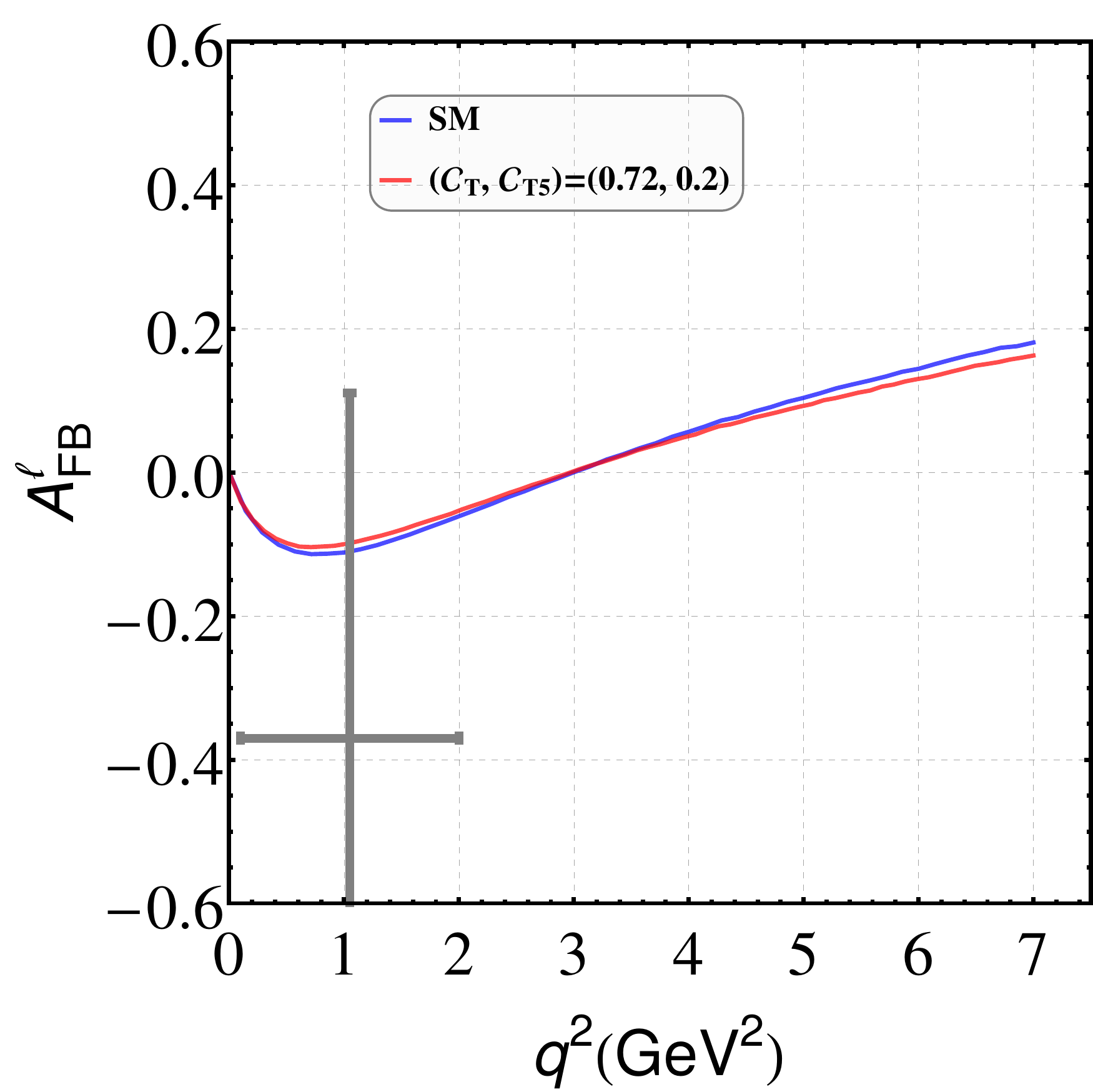}
		\includegraphics[scale=0.33]{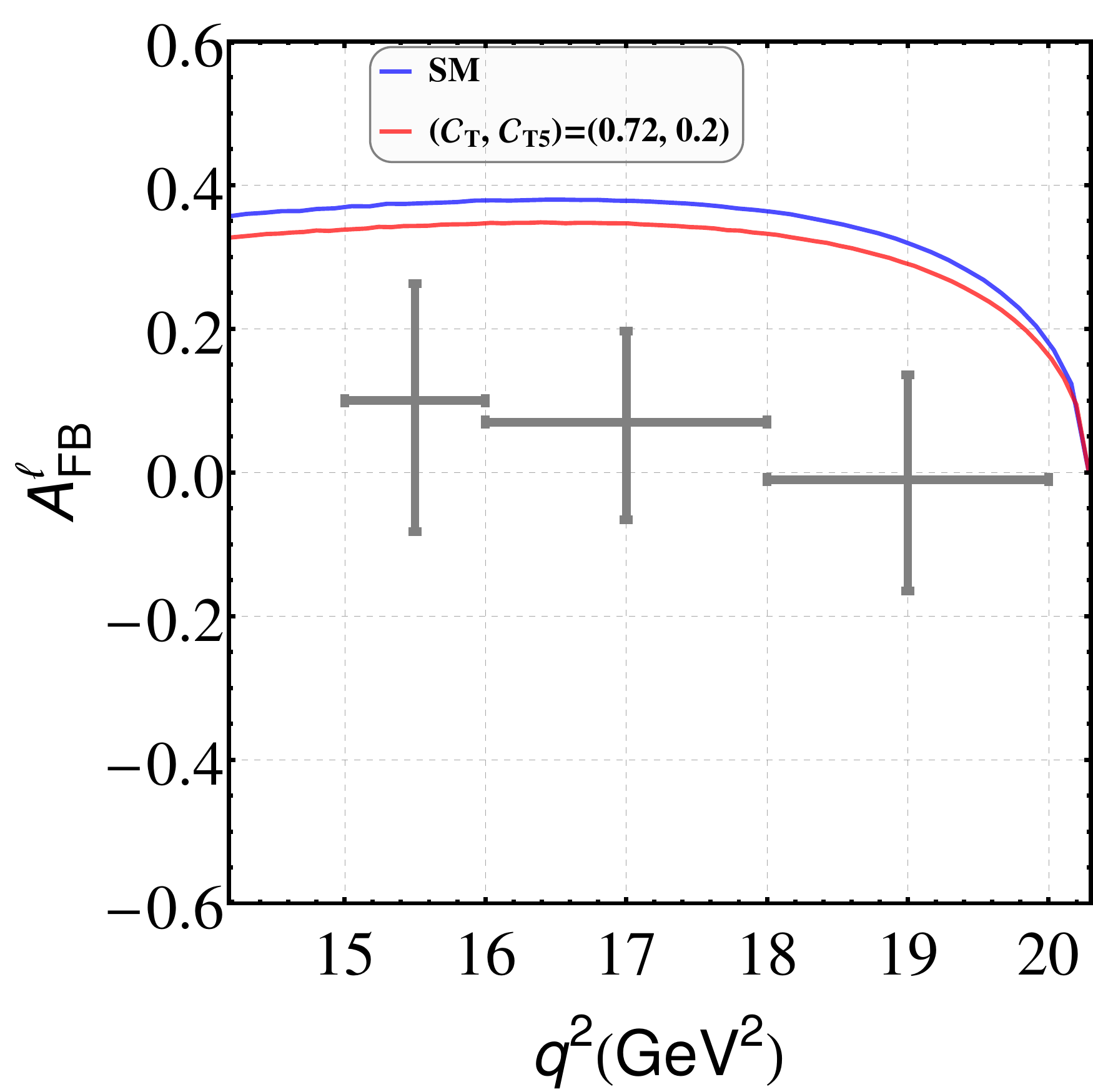}
		\caption{The $\Lambda_b \to \Lambda\mu^+\mu^-$ observables in the SM (blue) and for some new tensor couplings (shades of red). The points shown by cross marks indicate LHCb data \cite{Aaij:2015xza}.  \label{fig:T}}
	\end{center}
\end{figure}

The tensor couplings are constrained by the inclusive $\bar{B}\to X_s\mu^+\mu^-$ branching ratio. This is a theoretically clean mode for indirect searches of NP. The $q^2$-cut dependent branching fractions has been measured by Belle \cite{Iwasaki:2005sy} and BaBar \cite{Aubert:2004it,Lees:2013nxa} and the weighted average of the results read
\begin{eqnarray}\label{eq:XsllExp}
\mathcal{B}(\bar{B}\to X_s\mu^+\mu^-)^{[1,6]\rm GeV}_{\rm exp} &=& (1.58 \pm 0.37) \times 10^{-6}\,,\nn\\
\mathcal{B}(\bar{B}\to X_s\mu^+\mu^-)^{q^2 > 14.2,\rm GeV}_{\rm exp} &=& (0.48 \pm 0.10) \times 10^{-6}\, .
\end{eqnarray}
In the presence of the tensor couplings, the expression of $q^2$-integrated branching ratio following \cite{Fukae:1998qy} is
\begin{eqnarray}\label{eq:XsllTh}
\frac{d\mathcal{B}(\bar{B}\to X_s\mu^+\mu^-)}{dq^2} &=& \mathcal{B}(\bar{B}\to X_s\mu^+\mu^-)|_{\rm SM} + \frac{\mathcal{B}_0}{2m_b^8}  16M_9(q^2)(|\mC_T|^2 + |\mC_{T5}|^2) \, ,
\end{eqnarray}
where $\mathcal{B}(\bar{B}\to X_s\mu^+\mu^-)|_{\rm SM}$ is the branching ratio in the SM. The normalization factor is 
\begin{eqnarray}
\mathcal{B}_0 &=& \frac{3\alpha_e^2}{16\pi^2} \frac{|V_{tb}V_{ts}^\ast|^2}{|V_{cb}|^2} \frac{\mathcal{B}(B\to X_s\ell\nu)}{f(\hat{m}_c)\kappa(\hat{m}_c)}\, ,\nn\\
f(\hat{m}_c) &=& 1 - 8 \hat{m}_c^2 + \hat{m}_c^6 - \hat{m}_c^8 - 24 \hat{m}_c^4 \ln(\hat{m}_c)\, ,\quad \hat{m}_c = \frac{m_c}{m_b}\nn\\
\kappa(\hat{m}_c) &=& 1 - \frac{2\alpha_s(m_b)}{3\pi} \Bigg[\bigg( \pi^2 - \frac{31}{4} \Bigg)(1 - \hat{m}_c)^2 + \frac{3}{2}  \Bigg]\, ,\nn
\end{eqnarray}
where we use $\mathcal{B}(\bar{B}\to X_c\ell\nu) = 10.33\%$ \cite{Patrignani:2016xqp} and the quark masses are in the pole mass scheme. The experimental data (\ref{eq:XsllExp}) imply severe constraint on $\mC_{T,T5}$ couplings
\begin{equation}
\mC_T^2+\mC_{T5}^2 \lesssim 0.55\,, 
\end{equation}
and due to this the effects of tensor couplings on $\Lambda_b\to\Lambda\ell^+\ell^-$ are negligibly small as shown in Fig.~\ref{fig:T}.
%

\subsection{Combinations of VA, SP and T couplings}
\begin{figure}[h!]
	\begin{center}
		\includegraphics[scale=0.33]{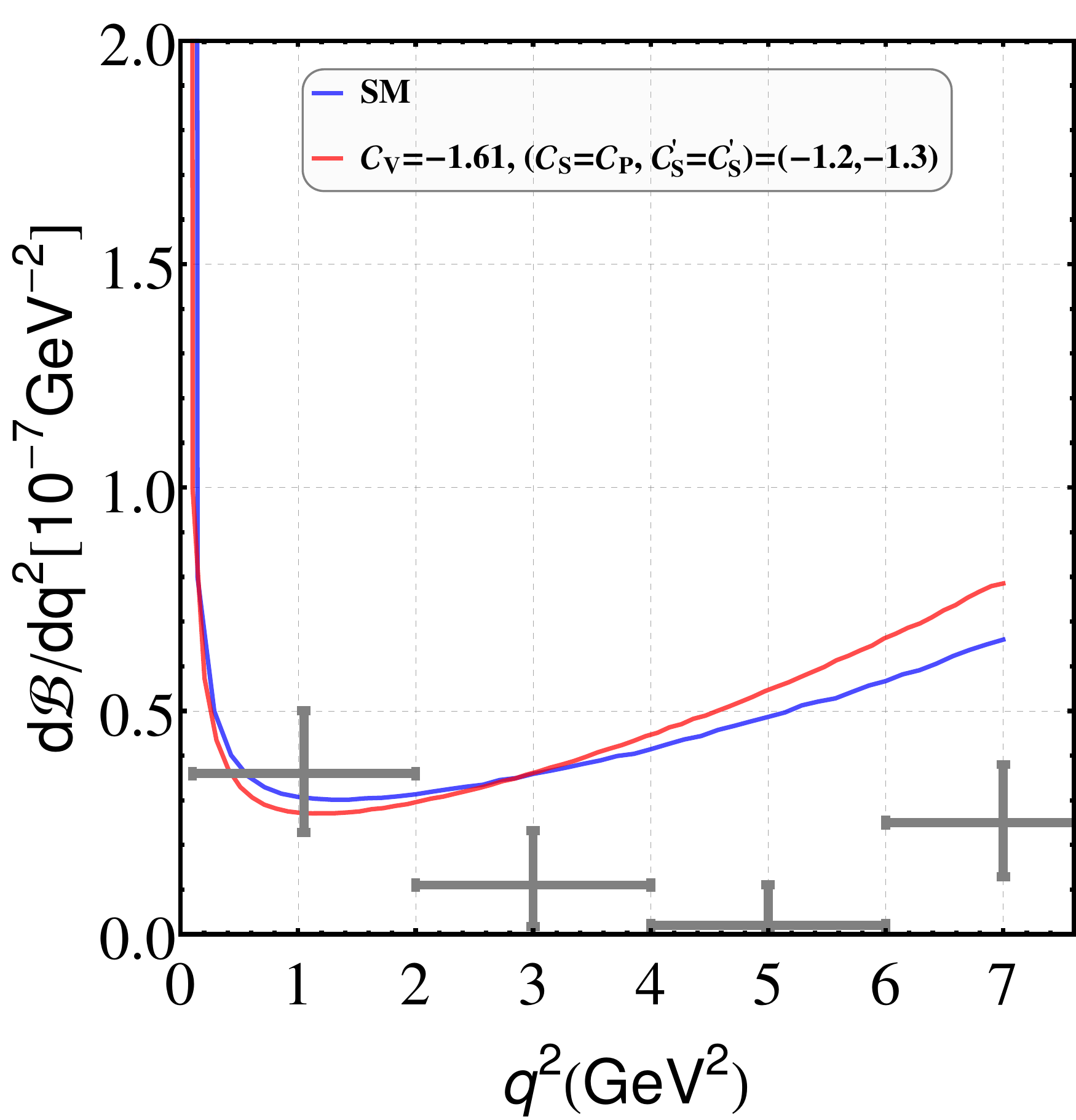}
		\includegraphics[scale=0.33]{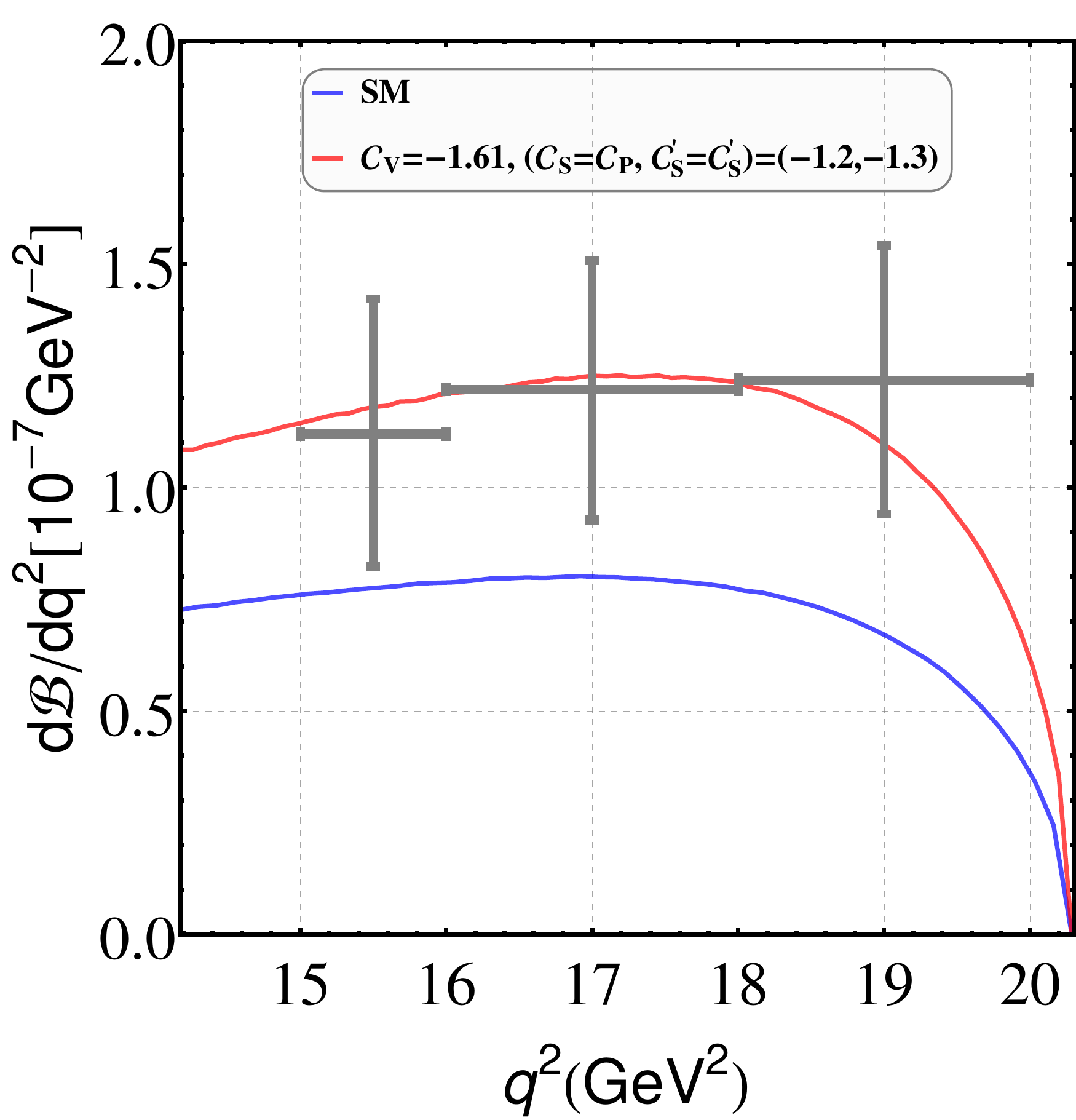}
		\includegraphics[scale=0.33]{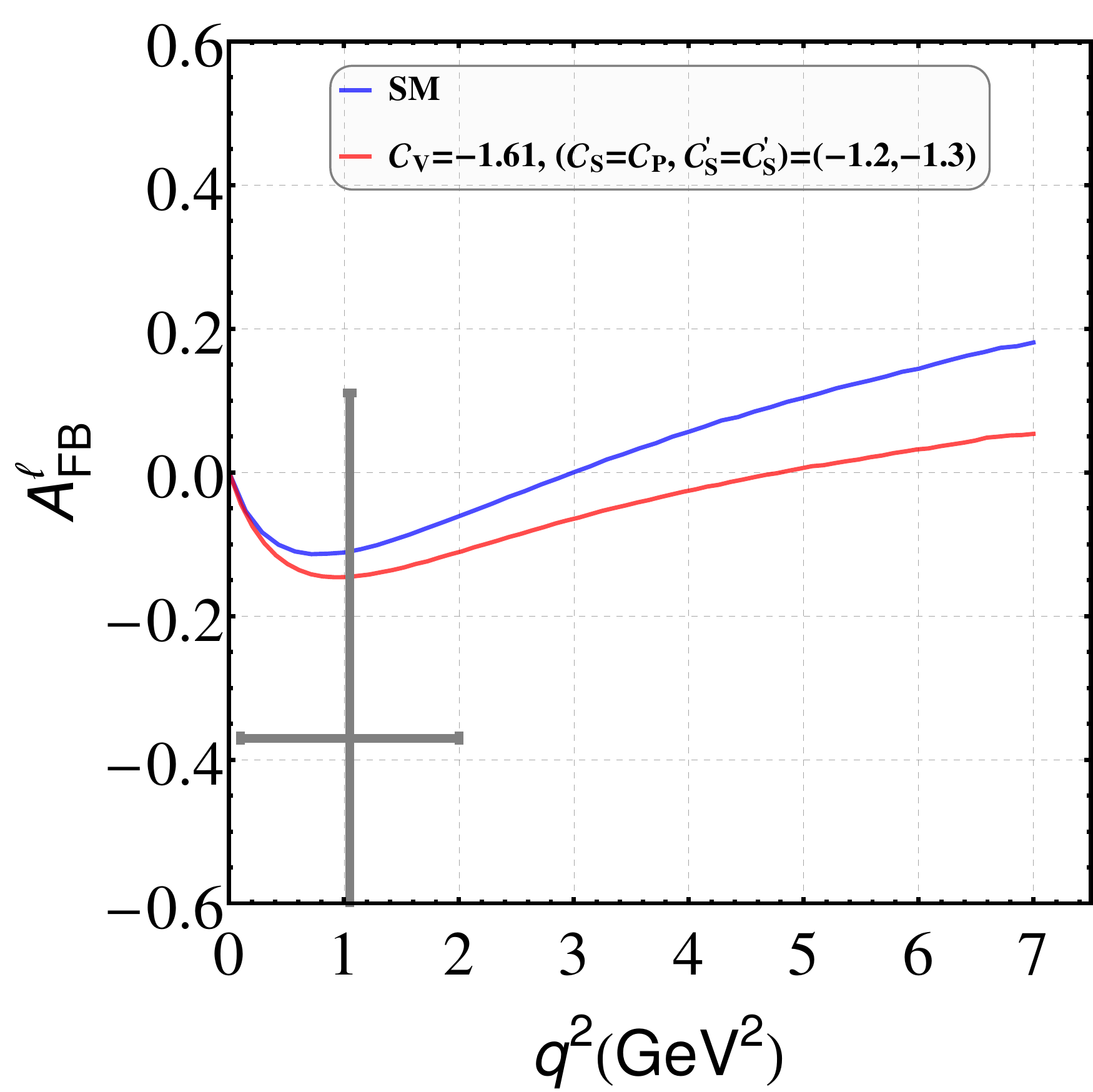}
		\includegraphics[scale=0.33]{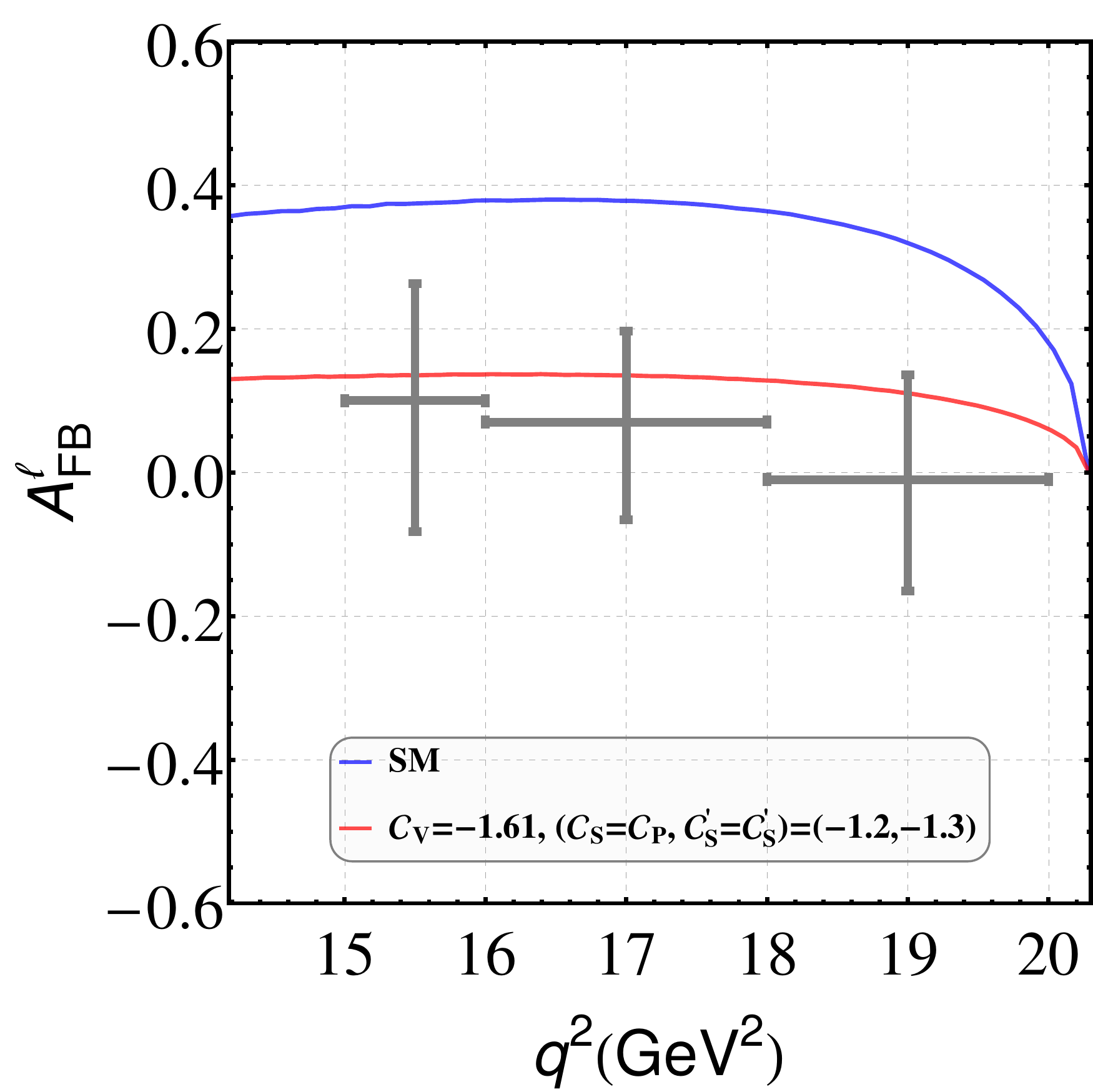}
		\caption{The $\Lambda_b \to \Lambda\ell\ell$ observables in the SM (blue) and in the simultaneous presence of VA and SP couplings(red). The points shown by cross marks indicate LHCb data \cite{Aaij:2015xza}.  \label{fig:VASP}}
	\end{center}
\end{figure}

After studying the patterns of the observables when VA, SP and T couplings are present separately, we study what happens when both VA and SP couplings are present. The terms $A^{\rm VA}$ and $A^{\rm SP}$ both contribute in this case.  We considered all the four scalar couplings $\mC_{S,P}^{(\prime)}$ simultaneously along with different combinations of VA couplings considered in Sec.~\ref{sec:VA}. Due to the interplay between VA and SP couplings, the resultant effects on branching ratio depend on the choice of values of the couplings. In general, if $\mC_{S,P}^{(\prime)}$ are large so that $A^{\rm SP}$ dominates over $A^{\rm VA}$, then branching ratio increases. Presence of both $\mC_{V,A}^{(\prime)}$ and $\mC_{S,P}^{(\prime)}$ has interesting effects on $A^\ell_{\rm FB}$. Since $A^{\rm SP}$ do not contribute to the numerator of $A^\ell_{\rm FB}$, the zero-crossings in the simultaneous presence of $\mC_{V,A}^{(\prime)}$ and $\mC_{S,P}^{(\prime)}$ are same as when only $\mC_{V,A}^{(\prime)}$ couplings are present. But the absolute value $|A^\ell_{FB}|$ is reduced compared to when only VA couplings are present and this is shown in Fig.~\ref{fig:VASP} (see also Fig.~\ref{fig:VA} to compare).

We also considered combinations of SP and T couplings. In this case in addition to $A^{\rm SP}$ and $A^{\rm T}$ the interference term $A^{\rm inter}$ contribute to the differential distribution. We find that due to stringent constraints on tensor couplings, the effects of combinations of SP and T couplings are almost identical to that when only SP couplings are present.

\section{Summary \label{sec:sum}}
In this paper we have investigated the effects of NP couplings on the $b\to s\mu^+\mu^-$ mediated rare decay $\Lambda_b\to\Lambda\mu^+\mu^-$. The NP considered are new VA operators, SP and T operators. We work in the helicity formalism and derive the expressions of helicity amplitudes with NP operators. We give the expressions of two fold differential distribution and study the observables $A^\ell_{\rm FB}$ and $F_L$ in the presence of the NP couplings. The hadronic matrix elements are defined in terms of so called helicity form factors. At the large recoil the form factors are taken from LCSR calculations while at the low recoil the form factors are taken from calculations in lattice QCD, and both are fitted to $z$-parametrization which give their $q^2$ dependence.

We sequentially discuss the implications of the NP couplings on differential branching ratio, $A^\ell_{\rm FB}$ and $F_L$. To understand the implications of the new VA couplings, we have followed the trends of these couplings suggested by the global fits to $b\to s\mu^+\mu^-$ data. We find that for most of the cases that are largely favored by the global fits, the branching ratio is reduced which is somewhat preferred by the LHCb data at low $q^2$ but is disfavored at large $q^2$. However, we have also identified a case for which the branching ratio increases.  In leptonic forward-backward asymmetry, the zero-crossings shift to the higher value of $q^2$ for most of the cases. We have however identified scenarios where it can shit to the lower value or remain SM like.

The scalars couplings are constrained together by the $\bar{B}_s\to\mu^+\mu^-$ and the inclusive $\bar{B}\to X_s\mu^+\mu^-$ branching ratios. We show that in the presence of only scalar couplings, the $\Lambda_b\to\Lambda\mu^+\mu^-$ branching ratio increases while the zero crossing of $A^\ell_{\rm FB}$ remain SM like. The tensor couplings however are severely constrained by the exclusive $\bar{B}\to X_s\mu^+\mu^-$ so that the effects on $\Lambda_b\to\Lambda\mu^+\mu^-$ are negligibly small. If VA and SP couplings are simultaneously present, the $A^\ell_{\rm FB}$ zero-crossing remain VA like, but $|A^\ell_{\rm FB}|$ is reduced compared to when only VA couplings are present.

\section*{Acknowledgements}
	The author is supported by the DST, Govt. of India under INSPIRE Faculty Fellowship.

\appendix

\section{SM Wilson coefficients \label{app:C910}}
The effective Wilson coefficient $\mathcal{C}_9^{\rm eff}$ read
\begin{equation}
\mathcal{C}_9^{\rm eff}(\mu) = \mathcal{C}_9^{\rm SM}(\mu) + Y^{\rm pert}(q^2) \, ,
\end{equation}
where $\mathcal{C}_9^{\rm SM}(\mu)$ indicates the value in the SM at the scale of $b$-quark mass, $\mu_b \approx \mathcal{O}(m_b)=4.8$ GeV \cite{Altmannshofer:2008dz}. The contributions from one-loop matrix elements of the four-quark operators \cite{Buras:1994dj,Misiak:1992bc} are contained in $Y^{\rm pert}(q^2)$ and is given by 
\begin{eqnarray}
Y^{\rm pert}(q^2) &=& h(m_c,q^2) \big( 3\mathcal{C}_1 + \mathcal{C}_2 + 3\mathcal{C}_3 + \mathcal{C}_4 + 3\mathcal{C}_5 + \mathcal{C}_6 \big)\, \nn\\
&-&\frac{1}{2} h(m_b,q^2) (4\mathcal{C}_3 + 4\mathcal{C}_4 + 3\mathcal{C}_5 + \mathcal{C}_6)- \frac{1}{2}h(0,q^2) (\mathcal{C}_3 + 3\mathcal{C}_4)\, \nn\\
&+& \frac{2}{9} (3\mathcal{C}_3 + \mathcal{C}_4 + 3\mathcal{C}_5 + \mathcal{C}_6)\, ,
\end{eqnarray}
where the loop functions read
\begin{eqnarray}
h(m_q, q^2) &=& -\frac{8}{9}\ln\frac{m_q}{m_b} + \frac{8}{27} + \frac{4}{9}x - \frac{2}{9}(2+x)|1-x|^{1/2} \left\{
\begin{array}{ll}
\ln\left| \frac{\sqrt{1-x} + 1}{\sqrt{1-x} - 1}\right| - i\pi, & x \equiv \frac{4 m_c^2}{ q^2} < 1,  \\ & \\ 2 \arctan \frac{1}{\sqrt{x-1}}, & x \equiv \frac {4 m_c^2}{ q^2} > 1,
\end{array}
\right. \\
h(0,q^2) & =& \frac{8}{27} - \frac{4}{9} \ln\frac{q^2}{m_b^2}  + \frac{4}{9} i\pi.
\end{eqnarray}
The quark masses appearing in the loop functions are in the pole mass scheme. All the values of SM Wilson coefficients including $\mC_{10}$ and $\mC_7^{\rm eff}$ are taken from \cite{Altmannshofer:2008dz}.

\section{Polarization conventions \label{app:pol}}
In the $\Lambda_b$ rest frame, the polarization four-vectors of the virtual gauge boson that decays to dilepton pair is 
\begin{equation}
\bar{\epsilon}^\mu(\pm) = \frac{1}{\sqrt{2}}(0, \pm 1, -i, 0)\, ,\quad
\bar{\epsilon}^\mu(0) = \frac{1}{\sqrt{q^2}}(|\textbf{q}|, 0, 0, -q^0)\, ,\quad
\bar{\epsilon}^\mu(t) = \frac{1}{\sqrt{q^2}}(q^0, 0, 0, -|\textbf{q}|)\, .
\end{equation} 
The polarization vectors satisfy the following orthonormality and completeness relations
\begin{equation}
\bar{\epsilon}^{\ast\mu}(n) \bar{\epsilon}_{\mu}(n^\prime) = g_{nn^\prime}\, ,\quad \sum_{n,n^\prime} \bar{\epsilon}^{\ast\mu}(n) \bar{\epsilon}^{\nu}(n^\prime) g_{nn^\prime} = g^{\mu\nu}\, ,\quad n, n^\prime = t, \pm 1, 0\, ,
\end{equation}
where $g_{n,n^\prime} =  diag(+1,-1,-1,-1)$ and our choice of the metric tensor is $g^{\mu\nu}=diag(1,-1,$ $-1,-1)$.
In the rest frame of the $\ell^+\ell^-$ system, the transversity polarizations of virtual gauge boson remain same where as the longitudinal and time like polarizations read $\bar{\epsilon}^\mu(0) = (0;0,0,-1)$ and $\bar{\epsilon}^\mu(t) = (1;0,0,0)$.

\section{$\Lambda_b \to \Lambda$ hadronic matrix elements \label{sec:hme}}
The $\Lambda_b \to \Lambda$ hadronic matrix elements are conveniently written in the helicty basis \cite{Feldmann:2011xf}. For the vector currents we have
\begin{align}\label{eq:VAhme1}
\langle \Lambda(k,s_k)|\bar{s}\gamma^\mu b |\Lambda(p,s_p)\rangle =& \bar{u}(k,s_k)\Bigg[f^V_t(q^2)(\mLb-\mL)\frac{q^\mu}{q^2}\nn\\ +& f^V_0(q^2) \frac{\mLb+\mL}{s_+} \{p^\mu + k^\mu  - \frac{q^\mu}{q^2}(\mmLb - \mmL) \} \nn\\ + & f^V_\perp(q^2) \{ \gamma^\mu - \frac{2\mL}{s_+}p^\mu - \frac{2\mLb}{s_+}k^\mu \} \Bigg]u(p,s_p)\, ,
\end{align}
where the variables $s_\pm$ are defined in Eq.~(\ref{eq:spm}). For the axial-vector currents we get
\begin{align}\label{eq:VAhme2}
\langle \Lambda(k,s_k)|\bar{s}\gamma^\mu\gamma_5 b |\Lambda(p,s_p)\rangle =& - \bar{u}(k,s_k) \gamma_5 \Bigg[ f_t^A(q^2) (\mLb + \mL) \frac{q^\mu}{q^2} \nn\\ +& f_0^A(q^2) \frac{\mLb - \mL}{s_-} \{p^\mu + k^\mu - \frac{q^\mu}{q^2} (\mmLb - \mmL) \} \nn\\ + & f_\perp^A(q^2) \{\gamma^\mu + \frac{2\mL}{s_-}p^\mu - \frac{2\mLb}{s_-}k^\mu \}  \Bigg] u(p,s_p)\, .
\end{align}
The matrix elements for the scalar and the pseudo-scalar currents can be obtained from Eqs.~(\ref{eq:VAhme1}) and (\ref{eq:VAhme2}) via the equations of motion
\begin{align}\label{eq:SPff1}
\langle \Lambda(k,s_k)|\bar{s} b |\Lambda(p,s_p)\rangle =& f^V_t(q^2) \frac{\mLb-\mL}{m_b} \bar{u}(k,s_k)u(p,s_p)\, ,\\
\label{eq:SPff2}
\langle \Lambda(k,s_k)|\bar{s} \gamma_5 b |\Lambda(p,s_p)\rangle =& f_t^A(q^2) \frac{\mLb + \mL}{m_b} \bar{u}(k,s_k) \gamma_5 u(p,s_p)\, ,
\end{align}
where we have neglected the mass of the strange quark in the denominator. For the dipole operators we get 
\begin{eqnarray}
\langle \Lambda |\bar{s}i q_\nu\sigma^{\mu\nu}b|\Lambda_b\rangle &=& -\bar{u}(k,s_k)\Bigg[ f^T_0(q^2) \frac{q^2}{s_+}\Bigg(p^\mu + k^\mu - \frac{q^\mu}{q^2}(\mmLb - \mmL) \Bigg) \,\nn\\ &+& f^T_\perp (\mLb+\mL) \Bigg( \gamma^\mu - \frac{2\mL}{s_+}p^\mu - \frac{2\mLb}{s_+}k^\mu \Bigg) \Bigg]u(p,s_p)\, ,
\end{eqnarray}
and 
\begin{eqnarray}
\langle \Lambda |\bar{s}i q_\nu\sigma^{\mu\nu}\gamma_5 b|\Lambda_b\rangle &=& -\bar{u}(k,s_k)\gamma_5 \Bigg[f^{T5}_0 \frac{q^2}{s_-} \Bigg( p^\mu + k^\mu - \frac{q^\mu}{q^2}(\mmLb - \mmL) \Bigg) \, \nn\\ &+& f^{T5}_\perp (\mLb - \mL) \Bigg( \gamma^\mu + \frac{2\mL}{s_-}p^\mu - \frac{2\mLb}{s_-}k^\mu \Bigg) \Bigg] u(p,s_p)\, .
\end{eqnarray}
The hadonic matrix elements for the tensor current is given by
\begin{eqnarray}
\langle \Lambda |\bar{s}i\sigma^{\mu\nu}b|\Lambda_b\rangle &=& \bar{u}(k,s_k)\bigg[ 2 f_0^{T}(q^2) \frac{p^\mu k^\nu - p^\nu k^\mu}{s_+} + f_\perp^{T}(q^2) \bigg( \frac{ m_{\Lambda_b} + m_{\Lambda} }{q^2}(q^\mu \gamma^\nu-q^\nu \gamma^\mu)\, \nn\\ &-& 2 (\frac{1}{q^2} + \frac{1}{s_+}) (p^\mu k^\nu - p^\nu k^\mu) \bigg) + f_0^{T5} \bigg( i\sigma^{\mu\nu} - \frac{2}{s_-}\big[m_{\Lambda_b}(k^\mu \gamma^\nu - k^\nu \gamma^\mu) \nn\\ &-& m_\Lambda (p^\mu \gamma^\nu - p^\nu \gamma^\mu) + p^\mu k^\nu - p^\nu k^\mu \big]   \bigg) + f_\perp^{T5}(q^2) \frac{ m_{\Lambda_b} - m_\Lambda }{q^2 s_-}\nn\\&\times&\bigg( (m_{\Lambda_b}^2 - m_\Lambda^2 - q^2 )(\gamma^\mu p^\nu - \gamma^\nu p^\mu) - ( m_{\Lambda_b}^2 - m_\Lambda^2 + q^2)(\gamma^\mu k^\nu - \gamma^\nu k^\mu)\, \nn\\ &+& 2(m_{\Lambda_b} - m_\Lambda )(p^\mu k^\nu - p^\nu k^\mu) \bigg) \bigg]u(p,s_p)\, .
\end{eqnarray}
%

\section{Spinor bilinears \label{app:spinor}}
To calculate the hadronic helicity amplitudes, we give the explicit expression of the spinor matrix elements. We follow the spinor representations given in \cite{Haber:1994pe} for these calculations. For (pseudo-)scalar operators we have the following non-vanishing elements
\begin{align}
&\bar{u}(k,\pm 1/2)u(p,\pm 1/2) = \sqrt{s_+}\, ,\\
&\bar{u}(k,\pm 1/2)u(p,\mp 1/2) = 0\, ,\\
&\bar{u}(k,\pm 1/2)\gamma_5 u(p,\pm 1/2) = \mp \sqrt{s_-}\, ,\\
&\bar{u}(k,\pm 1/2)\gamma_5 u(p,\mp 1/2) = 0\, ,
\end{align}
and for (axial-)vector operators we get
\begin{align}
&\bar{u}(k,\pm 1/2) \gamma^\mu u(p,\pm 1/2) = (\sqrt{s_+}, 0, 0, \sqrt{s_-})\, ,\\
&\bar{u}(k,\pm 1/2) \gamma^\mu u(p,\mp 1/2) = \sqrt{2 s_-} \epsilon^\mu(\pm)\, ,\\
&\bar{u}(k,\pm 1/2) \gamma^\mu\gamma_5 u(p,\pm 1/2) = \pm(\sqrt{s_-}, 0, 0, \sqrt{s_+})\, ,\\
\label{eq:pmA}
&\bar{u}(k,\pm 1/2) \gamma^\mu\gamma_5 u(p,\mp 1/2) = \pm \sqrt{2s_+} \epsilon^\mu(\pm)\, .
\end{align}
The non-vanishing elements of the tensor operators are
\begin{align}
&\bar{u}(k,\pm1/2) \begin{Bmatrix} \sigma^{14}\\ \sigma^{23} \end{Bmatrix} u(p,\pm1/2)  = \begin{Bmatrix} -i\sqrt{s_-} \\ \pm\sqrt{s_+} \end{Bmatrix}\, ,\\
&\bar{u}(k,\pm 1/2) \begin{Bmatrix} \sigma^{12} \\ \sigma^{13}\\ \sigma^{24}\\ \sigma^{34} \end{Bmatrix} u(p,\mp 1/2)  = \begin{Bmatrix} \mp i\sqrt{s_-} \\ -\sqrt{s_-} \\ \pm i \sqrt{s_+} \\ \sqrt{s_+} \end{Bmatrix}\, ,
\end{align}
where $\bar{u}(k,s_k)\sigma^{\mu\nu}u(p,s_p) = - \bar{u}(k,s_k)\sigma^{\nu\mu}u(p,s_p)$.


\section{Spinors in dilepton rest frame \label{sec:llRF}}
To calculate the lepton helicity amplitudes, we work in the dilepton rest frame. The angle $\theta_\ell$ is defined as the angle made the negatively charged lepton $\ell^-$ with the $\Lambda_b$ in the dilepton rest frame. The four momentum of the charged leptons $\ell^-$ and $\ell^+$ (here we use the subscript $2\ell$ to denote this frame) respectively are
\begin{align}
& q_-^\mu \Big|_{2\ell} = (E_\ell, |q_{2\ell}|\sin\theta_\ell, 0, |q_{2\ell}|\cos\theta_\ell)\, ,\\
&q_+^\mu \Big|_{2\ell} = (E_\ell, -|q_{2\ell}|\sin\theta_\ell, 0, -|q_{2\ell}|\cos\theta_\ell)\, ,
\end{align}
with 
\begin{equation}
|q_{2l}| = \frac{\beta_\ell}{2}\sqrt{q^2}\, ,\quad\quad E_\ell = \frac{\sqrt{q^2}}{2}\, ,\quad\quad \beta_\ell = \sqrt{1-\frac{4m_\ell^2}{q^2}}\, .
\end{equation}
The explicit expressions of the amplitudes in Eqs.~(\ref{eq:Ldef1})-(\ref{eq:Ldef2}) are obtained by contractions of the currents
\begin{equation}
\bar{u}_{\ell^-} \{1,\gamma_\mu, \sigma_{\mu\nu} \} (1\mp\gamma_5) v_{\ell^+}\, ,
\end{equation}
with $1, \bar{\epsilon}^\mu(\lambda), -i\bar{\epsilon}^\mu(\lambda)\bar{\epsilon}^\nu(\lambda^\prime)$ respectively.
Following \cite{Haber:1994pe} we give the explicit expressions of the spinors in this frame are
\begin{align}
& u_{\ell^-}(\lambda) = 
\begin{pmatrix}
\sqrt{E_\ell+m_\ell} \chi^u_\lambda  \\ 2 \lambda \sqrt{E_\ell-m_\ell} \chi^u_\lambda
\end{pmatrix}\, ,
\quad \chi^u_{+\frac{1}{2}} = \begin{pmatrix} \cos\frac{\theta_\ell}{2} \\ \sin\frac{\theta_\ell}{2} \end{pmatrix}\, ,
\quad \chi^u_{-\frac{1}{2}} = \begin{pmatrix} -\sin\frac{\theta_\ell}{2} \\ \cos\frac{\theta_\ell}{2} \end{pmatrix}\, ,\\
& v_{\ell^+}(\lambda) = 
\begin{pmatrix}
\sqrt{E_\ell-m_\ell} \chi^v_{-\lambda}  \\ -2 \lambda \sqrt{E_\ell+m_\ell} \chi^v_{-\lambda}
\end{pmatrix}\, ,
\quad \chi^v_{+\frac{1}{2}} = \begin{pmatrix} \sin\frac{\theta_\ell}{2} \\ -\cos\frac{\theta_\ell}{2} \end{pmatrix}\, ,
\quad \chi^v_{-\frac{1}{2}} = \begin{pmatrix} \cos\frac{\theta_\ell}{2} \\ \sin\frac{\theta_\ell}{2} \end{pmatrix}\, .
\end{align}

\section{Numerical inputs \label{sec:inputs}}

In the following table we collect the numerical values of the inputs used in the paper.

\begin{longtable}{cr|cr}
	\hline\hline
	inputs  & values  & inputs  & values   \\ 
	\hline 
	$ \alpha_e(m_b) $  & $1/127.925(16)$ \cite{Patrignani:2016xqp} & $|V_{tb}V_{ts}^\ast|$  & $0.0401 \pm 0.0010$ \cite{Bona:2006ah}  \\ 
	$m_c(\overline{\text{MS}})$ & 1.28 GeV \cite{Patrignani:2016xqp} & $\mLb$ & 5.619 GeV \cite{Patrignani:2016xqp} \\ 
	$\mu_b$ & $4.8$ GeV \cite{Altmannshofer:2008dz} & $\mL$ & $1.115$ GeV \cite{Patrignani:2016xqp} \\
	$m_b(\overline{\text{MS}})$ & $4.2$ GeV \cite{Altmannshofer:2008dz} & $\tau_{\Lambda_b}$ & $(1.470\pm 0.010)\times 10^{-12} s$ \cite{Patrignani:2016xqp} \\  
	$m_{b}({\rm pole})$ & $4.8$ GeV \cite{Altmannshofer:2008dz}  & $m_{B^0}$ & $5.279$ GeV \cite{Detmold:2016pkz} \\ 
	$\alpha_s(m_b)$ & $0.214$ \cite{Altmannshofer:2008dz}  & $m_K$ & $0.494$ GeV \cite{Detmold:2016pkz} \\ 
	$|V_{cb}|$ & $(4.15 \pm 0.07)\times 10^{-2}$ \cite{Bona:2006ah} & &\\
	\hline\caption{List of inputs and their values. \label{tab:inputs}}
\end{longtable}

\end{document}